\documentclass[11pt]{article}

\usepackage[utf8]{inputenc}
\usepackage[english]{babel}
\usepackage{amsmath}
\usepackage{amsfonts}
\usepackage{amssymb}
\usepackage{amsthm}
\usepackage{graphicx}
\usepackage{xcolor}

\usepackage{lineno}
\usepackage{subcaption}
\usepackage{tikz}
\usetikzlibrary{arrows.meta}
\usetikzlibrary{matrix, arrows, positioning, calc, shapes}

\usepackage[sort&compress, numbers]{natbib} 

\graphicspath{{figures/}}

\theoremstyle{plain}
\newtheorem{theorem}{Theorem}[section]
\newtheorem{example}[theorem]{Example}
\newtheorem{remark}[theorem]{Remark}
\theoremstyle{definition}
\newtheorem{definition}[theorem]{Definition}

\newcommand{\wendy}{{\bf WENDyS}}

\newcommand{\minnode}{\mathsf{MN}}

\theoremstyle{plain}  

\theoremstyle{definition}  

\theoremstyle{remark}  

\newcommand{\setof}[1]{\left\{ {#1}\right\}}

\newcommand{\R}{{\mathbb{R}}}

\newcommand{\cF}{{\mathcal F}}

\newcommand{\cX}{{\mathcal X}}

\newcommand{\sE}{{\mathsf E}}

\newcommand{\sH}{{\mathsf H}}

\newcommand{\sL}{{\mathsf L}}

\newcommand{\sMG}{{\mathsf{ MG}}}

\DeclareMathOperator{\sgn}{sgn}

\definecolor{gray85}{gray}{0.85} 
\definecolor{gray8}{gray}{0.8} 
\definecolor{gray7}{gray}{0.7} 
\definecolor{gray6}{gray}{0.6} 
\definecolor{gray5}{gray}{0.5} 
\definecolor{gray4}{gray}{0.4} 
\definecolor{gray35}{gray}{0.35} 

\def\corcommstyle{\bf\small\tt}

\def\corrl #1<<#2||#3>>{
\if\visiblecomments y
  \begin{quote} {\corcommstyle $<<$COMMENT$>>$ {\color{red}#1\marginpar{!!}}\\$<<$OLD$<<$} \end{quote}

{\color{red} 
 #2
 }

  \begin{quote} {\corcommstyle ==NEW== } \end{quote}
   \noindent\hrulefill
 
\vspace{-10pt} 
 
 \noindent\hrulefill
 
 \vspace{-10pt} 
 
 \noindent\dotfill
 
  #3
  
   \noindent\dotfill 

\vspace{-10pt} 
 
 \noindent\hrulefill
 
 \vspace{-10pt} 
 
 \noindent\hrulefill
  \begin{quote} {\corcommstyle $>>$END$>>$ } \end{quote}
 \else
  #3
 \fi
}

\long\def\longcorrl #1<<#2||#3>>{
\if\visiblecomments y
  \begin{quote} {\corcommstyle $<<$COMMENT$>>$ {\color{red}#1\marginpar{!!}}\\$<<$OLD$<<$} \end{quote}
 
 {\color{red}

  #2
  
  }
  
  \begin{quote} {\corcommstyle ==NEW== } \end{quote}
  
    \noindent\hrulefill
 
\vspace{-10pt} 
 
 \noindent\hrulefill
 
 \vspace{-10pt} 
 
 \noindent\dotfill
 
  #3
  
   \noindent\dotfill 

\vspace{-10pt} 
 
 \noindent\hrulefill
 
 \vspace{-10pt} 
 
 \noindent\hrulefill
  \begin{quote} {\corcommstyle $>>$END$>>$ } \end{quote}
 \else
  #3
 \fi
}

\def\mlabel #1
{
  \if\visiblecomments y
     \marginpar[\flushright \bf \footnotesize #1]{\bf \footnotesize #1}
  \fi
}

\def\flabel #1
{
  \if\visiblecomments y
       \hbox{\bf\footnotesize #1}
  \fi
}

\def\corrq #1<<#2>>{
\if\visiblecomments y
  \begin{quote} {\corcommstyle $<<$COMMENT$>>$ {\color{red}#1}\marginpar{!!}\\$<<$BEG$<<$} \end{quote}
  \noindent\hrulefill
 
\vspace{-10pt} 
 
 \noindent\hrulefill
 
 \vspace{-10pt} 
 
 \noindent\dotfill

  #2
 
  \noindent\dotfill 

\vspace{-10pt} 
 
 \noindent\hrulefill
 
 \vspace{-10pt} 
 
 \noindent\hrulefill 
  \begin{quote} {\corcommstyle $>>$END$>>$ } \end{quote}
 \else
  #2
 \fi
}

\long\def\longcorrq #1<<#2>>{
\if\visiblecomments y
  \begin{quote} {\corcommstyle $<<$COMMENT$>>$ #1\marginpar{!!}\\$<<$BEG$<<$} \end{quote}
  \noindent\hrulefill
 
\vspace{-10pt} 
 
 \noindent\hrulefill
 
 \vspace{-10pt} 
 
 \noindent\dotfill

  #2

  \noindent\dotfill 

\vspace{-10pt} 
 
 \noindent\hrulefill
 
 \vspace{-10pt} 
 
 \noindent\hrulefill 
  \begin{quote} {\corcommstyle $>>$END$>>$ } \end{quote}
 \else
  #2
 \fi
}

\def\corrc #1<<>>{
\if\visiblecomments y
  \begin{quote} {\corcommstyle $<<$COMMENT$>>$ \color{red} #1\marginpar{!!}} \end{quote}
\fi
}

\def\corre #1<<#2||#3>>{
\if\visiblecomments y
  #3\marginpar{\corcommstyle #1}
 \else
  #3
 \fi
}

\long\def\longcorre #1<<#2||#3>>{
\if\visiblecomments y
  #3\marginpar{\corcommstyle #1}
 \else
  #3
 \fi
}

\def\corrs #1<<#2||#3>>{
\if\visiblecomments y
  #3\marginpar{\corcommstyle #2 $\rightarrow$ #3\\ #1}
 \else
  #3
 \fi
}

\def\corro #1<<#2||#3>>{
#2}

\def\corrn #1<<#2||#3>>{
#3}

\long\def\longcorro #1<<#2||#3>>{
#2}

\long\def\longcorrn #1<<#2||#3>>{
#3}

\long\def\underconstruction #1<<<#2>>>{
\if\visiblecomments y
  \begin{quote} {\corcommstyle $<<$UNDER CONSTRUCTION - BEGIN$>>$ #1\marginpar{!!}} \end{quote}
  #2
  \begin{quote} {\corcommstyle $>>$UNDER CONSTRUCTION - END$>>$ } \end{quote}
 \else
 \fi
}

\def\showcomments{
  \let\visiblecomments y
}

\def\hidecomments{
  \let\visiblecomments n
}

\showcomments

\title{Inferring Long-term Dynamics of Ecological Communities Using Combinatorics}

\author{William S. Cuello \thanks{Information Systems $\&$ Modeling, Los Alamos National Laboratory, Los Alamos, NM ({wscuello@lanl.gov})} 
\and
Marcio Gameiro{\thanks{Department of Mathematics, Rutgers University, Piscataway, NJ. ({gameiro@math.rutgers.edu}; {mischaikow@rutgers.edu})} 
\thanks{Instituto de Ci\^{e}ncias Matem\'{a}ticas e de Computa\c{c}\~{a}o, Universidade de S\~{a}o Paulo, S\~{a}o Carlos, S\~{a}o Paulo, Brazil.}}
\and Juan A. Bonachela{\thanks{Department of Ecology, Evolution, and Natural Resources, Rutgers University, New Brunswick, NJ. ({juan.bonachela@rutgers.edu})}}
\and Konstantin Mischaikow{\footnotemark[2]}
}

\date{}

\begin{document}

\maketitle

\begin{abstract}
In an increasingly changing world, predicting the fate of species across the globe has become a major concern. Understanding how the population dynamics of various species and communities will unfold requires predictive tools that experimental data alone can not capture. Here, we introduce our combinatorial framework, Widespread Ecological Networks and their Dynamical Signatures ($\wendy$) which, using data on the relative strengths of interactions and growth rates within a community of species predicts all possible long-term outcomes of the community.  To this end, $\wendy$ partitions the multidimensional parameter space (formed by the strengths of interactions and growth rates) into a finite number of regions, each corresponding to a unique set of coarse population dynamics. Thus, $\wendy$ ultimately creates a library of all possible outcomes for the community. On the one hand, our framework avoids the typical ``parameter sweeps'' that have become ubiquitous across other forms of mathematical modeling, which can be computationally expensive for ecologically realistic models and examples. On the other hand, $\wendy$ opens the opportunity for interdisciplinary teams to use standard experimental data (i.e., strengths of interactions and growth rates) to filter down the possible end states of a community. To demonstrate the latter, here we present a case study from the Indonesian Coral Reef. We analyze how different interactions between anemone and anemonefish species lead to alternative stable states for the coral reef community, and how competition can increase the chance of exclusion for one or more species. $\wendy$, thus, can be used to anticipate ecological outcomes and test the effectiveness of management (e.g., conservation) strategies.
\end{abstract}


\section{Significance Statement}

We present a mathematical framework ($\wendy$) that takes as an input a network of species and their interactions, and outputs all possible long-term outcomes of the community's population dynamics. By viewing population dynamics via a coarse combinatorial lens, we are able to find all long-term outcomes of the community as a function of the relative magnitudes of their interaction strengths and growth rates. The determination of all possible outcomes by using $\wendy$ with (even sparse) simulated or real data enables the quantification of how probable they are, and study how that probability changes across a variety of scenarios. Our framework can thus inform management decisions by providing an assessment of the risks (e.g., extinction probability) associated with the implementation of strategies that affect single components or the community as a whole.


\section{Introduction }

In an increasingly changing world, predicting the fate of species across the globe has become a major concern \citep{kimball2010contemporary,crowley2019predicting}.
To this end, research has focused on the interaction between species and predicting their long-term outcome as a community, oftentimes considering environmental feedback and effects such as climate change and habitat fragmentation \cite{kondoh2003foraging, gonzalez2011disentangled}. Given the spatio-temporal scales involved, however, empirical work aiming to gather evidence to study long-term community dynamics is labor and time intensive \citep{schimel2015big}. Thus, ecological modeling offers an unrivaled opportunity to this end.

Theoretical ecologists have traditionally used mathematical tools such as ordinary differential equations (ODEs) to model the dynamics of the interactions between populations of species and with their environment \citep{upadhyay2010dynamical,hastings1991chaos,may2019stability}. However, finding the long-term solutions to such ODEs becomes increasingly complex as more species and nonlinear interactions are considered in the system because each additional species introduces more parameters and increases the dimensionality of the phase space.  As a consequence, there is also great potential for perturbations such as environmental fluctuations to lead to qualitatively different population trajectories (i.e., bifurcations) \citep{zhang2019global,strogatz2018nonlinear, troost2007bifurcation}. Therefore, it becomes less feasible, even computationally, to explore the long-term behavior of realistic communities. Ideally, experimental data would be used to constrain the parameter space. However, experiments that can procure a sufficient amount of data are time-consuming and expensive, and thus typically limited \citep{reichman2011challenges}. Therefore, to understand and predict the long-term dynamics of a community, it is paramount to find a theoretical framework that overcomes the limitations associated with this high dimensionality, thus enabling the exploration of the vast number of qualitatively different community trajectories that can arise.

A framework used in the past to study community dynamics is topological network analysis. Directed graphs are used to model species (nodes) and their interactions (links and edges) \citep{proulx2005network,landi2018complexity}. This mathematical framework has been used extensively to address the role that nestedness, modularity, and diversity play in the observed configurations of trophic and mutualistic networks \citep{hui2019invade, fortuna2010nestedness, bascompte2006asymmetric}. A main debate has been whether certain configurations of interactions stabilize a community \citep{may1972will,may1973stability,goodman1975theory}, although the topic remains controversial due to the fluid nature of the definition of stability \citep{landi2018complexity}.  ODE theory lends a clear, mathematical lens through which to view stability, but cannot be applied to network analyses without suffering from the same aforementioned bottleneck of large systems and many coefficients when exploring the associated non-linear population dynamics \citep{stouffer2010understanding}. All this together reveals the need for a mathematical framework able to merge the powerful network approach with the robustness of the ODE toolbox and definitions. Here, we aim to fill this gap.

We present the Widespread Ecological Networks and their Dynamical Signatures ($\wendy$) computational framework which, using data on the presence and strength of interactions between species, provides a library of all possible long-term and coarse population dynamics. $\wendy$ is a wrapper to the Dynamic Signatures Generated by Regulatory Networks (DSGRN) software, which is based on a combinatorial and topological framework for nonlinear dynamics that has been successfully employed in the context of gene regulatory networks \citep{cummins2016combinatorial}.
To ensure ecological relevance, $\wendy$ both restricts and expands DSGRN capabilities, thereby allowing for relevant forms of interaction between species, reproductive growth of populations, and the identification of extinction.
$\wendy$ takes the information of a community as an input, more specifically a graph of nodes (species) and directed edges (interactions), and outputs a finite combinatorial characterization of coarse, qualitative population dynamics that can arise as a function of relative magnitudes of species’ growth rates and strengths of interactions between species.

To introduce the mathematical concepts underlying $\wendy$, we first study a single species theoretical example with intraspecific competition.
We then show how these ideas extend to higher dimensions by analyzing a classic theoretical two-species predator-prey system.
Using these concepts, we then consider a real-world mutualistic community of anemonefish and anemone, with which we illustrate how experimental data can be used to construct a graph of species and interactions and use $\wendy$ to produce a summary of possible long-term population dynamics that can arise for the community. 
$\wendy$ can identify possible long-term outcomes of larger communities (e.g., a coral reef hosting anemonefish and anemone species), bypassing the high-dimensional challenges associated with classical mathematical models. Our framework can thus inform management decisions by providing an assessment of the different outcomes and risks expected from the implementation of strategies that affect single components or the community as a whole.

\section{Widespread Ecological Networks and their Dynamical Signatures}

Our starting point is a network (graph) consisting of $k$ species (nodes) and their interactions (directed edges). The interactions are visually represented on the graph by either a pointed arrow $\rightarrow$ or a blunted arrow $\dashv$ (interaction with the source species benefits or detriments the target species, respectively; see examples in Figure~\ref{fig:1D_network_signs}A and Figure~\ref{fig:2species}A). $\wendy$ translates this graph into coarse population dynamics in the nonnegative space of densities $\mathbb{R}^k_+$.
Unlike ODE models, which view $\mathbb{R}^k_+$ as a density space consisting of an (uncountably) infinite number of possible density values, our framework codifies $\mathbb{R}^k_+$ via a finite cell complex $\cX$ \citep{lefschetz}.
This cell complex is dimension-dependent and, roughly speaking, is a collection of intervals and their boundaries (endpoints) if $k=1$ (see Figure~\ref{fig:1D_Morse}A), rectangles and their boundaries (sides) if $k=2$ (see Figure~\ref{fig:2species}B-C), or hyperrectangles and their boundaries if $k > 2$. We refer to these cells as \emph{density regions}.

Importantly, given the configuration of species and their interactions, $\wendy$ establishes equations for an explicit finite decomposition of the parameter space formed by the species’ growth rates and strength of interactions (see the Appendix for details). We organize this decomposition via a \emph{parameter graph}, where vertices represent regions of the parameter space and an edge between vertices indicates that the two parameter regions are neighbors (share a codimension-1 hypersurface). A fundamental property of the parameter graph is that the dynamics (as characterized by our approach) is constant over each parameter region.

For a given region of the parameter space and along the boundaries of density regions, the population of at least one species strictly increases or decreases.
From these increases and decreases, we derive a combinatorial model for the population dynamics, called a \emph{state transition graph} (STG). The STG provides information as to whether it is possible for a community whose state is in one cell of the density space $\R^k_+$ to transition to another neighboring cell.
Our characterization of the global dynamics takes the form of a \emph{Morse graph} (MG), described in greater detail below. 
The essential information is that the nodes of the MG correspond to recurrence in the STG and the directed edges of the MG indicate the direction of nonrecurrent dynamics.
Thus, the minimal nodes of an MG, i.e., nodes that have no out edges, represent the long-term asymptotic behavior of the community's population, and hence the goal of the framework is to identify minimal nodes of MGs over all regions of parameter space for the selected examples.  
If a MG has multiple minimal nodes, then the system exhibits multistability and thus there are
corresponding locally attracting density regions.

\subsection{Simplest example: single species with intraspecific competition}
\label{sec:intuition}

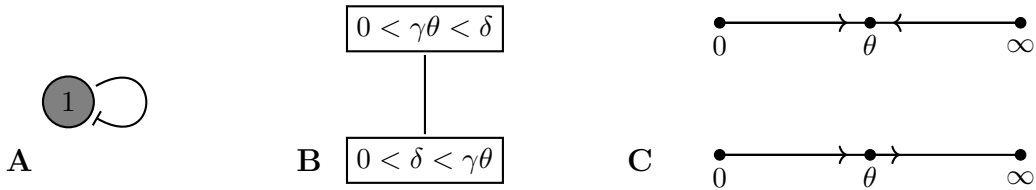
\begin{figure}[!htbp]
\centering
\begin{picture}(400,100)

\put(-10,20){{\bf \large A}}
\put(0,25){
\begin{tikzpicture}[thick, scale=0.9]
\node[circle, fill = gray, draw] (1) at (0,0) {1};
\draw[-|, shorten <= 2pt, shorten >= 2pt] (1) to [out=30, in=-30, looseness=8] (1);
\end{tikzpicture}
}

\put(100,20){{\bf \large B}}
\put(115,15){
\begin{tikzpicture}
    \node[rectangle, draw, thick] (2) at (0,0) {$0 < \gamma \theta < \delta$};
    \node[rectangle, draw, thick] (3) at (0,-1.75) {$0 < \delta < \gamma \theta $};
    \draw[-, shorten <= 1pt, shorten > =1pt, thick] (2) to [out = -90, in = 90, looseness = 2] (3);
\end{tikzpicture}
}

\put(225,20){{\bf \large C}}
\put(250,60){
\begin{tikzpicture}
\draw[thick, black] (0,0)--(4,0);
\filldraw[black] (0,0) circle (2pt);
\filldraw[black] (2,0) circle (2pt);
\filldraw[black] (4,0) circle (2pt);
\draw[->,thick] (0,0) -- (1.7,0);
\draw[->,thick] (4.0,0) -- (2.3,0);
\node at (2,-0.3) {$\theta$};
\node at (0,-0.3) {$0$}; 
\node at (4,-0.3) {$\infty$};
\end{tikzpicture}
}

\put(250,10){
\begin{tikzpicture}
\draw[thick, black] (0,0)--(4,0);
\filldraw[black] (0,0) circle (2pt);
\filldraw[black] (2,0) circle (2pt);
\filldraw[black] (4,0) circle (2pt);
\draw[->,thick] (0,0) -- (1.7,0);
\draw[->,thick] (2.0,0) -- (2.4,0);
\node at (2,-0.3) {$\theta$};
\node at (0,-0.3) {$0$}; 
\node at (4,-0.3) {$\infty$};
\end{tikzpicture}
}
\end{picture}
\caption{ \footnotesize {\bf Panel A}: Graph representation (a single node with a self-looping blunt edge) of a single species experiencing intraspecific competition. {\bf Panel B}: The parameter graph, where each vertex is a relative ordering of parameters that determine the rate of change $r$ of the population; the edge between them indicates that the relative orderings differ by an exchange of one inequality (mathematically, share a codimension-$1$ hypersurface). {\bf Panel C; top}: Carrying-capacity-like dynamics when intraspecific competition overcomes the background growth rate of the population (i.e., in parameter region $0 < \gamma\theta < \delta$). {\bf Panel C; bottom}: Potentially unbounded growth when the background growth rate of the population exceeds intraspecific competition (i.e., in parameter region $0 < \delta < \gamma\theta $).}
\label{fig:1D_network_signs}
\end{figure}

To illustrate how our combinatorial framework works, we present a model of single species population dynamics (i.e., $k=1$) and describe our visual representation of this model.
We assume that the species has a positive net growth rate $\gamma$.
The dynamics can thus be represented via a graph with one node (the species) that is colored grey to indicate a positive net growth rate (see Figure~\ref{fig:1D_network_signs}A).
We assume that the species' population experiences intraspecific competition at a rate $\delta > 0$, which becomes noticeable only if the population density surpasses a threshold $\theta > 0$.
This is visually indicated by a self-looping blunted edge (see Figure~\ref{fig:1D_network_signs}A). Note that  this model has three parameters, $\gamma$, $\delta$, and $\theta$, and thus the associated parameter space is $\R^3_+$.

Next, we build a combinatorial model of the dynamics in two steps.
First we identify, as a function of the parameters, whether the population density is increasing and/or decreasing ``at" the threshold $\theta$.
To be more precise, we decompose $\mathbb{R}_+$ into two regions: $[0, \theta]$ and $[\theta, \infty)$.
The rate of change $r$ of the population density $x$ is given by
\begin{equation}
\label{eq:rate_change}
r(x) = \gamma x -
\begin{cases}
0, & \text{if $x \in [0, \theta)$} \\
\delta, & \text{if $x \in (\theta,\infty)$}
\end{cases}    
\end{equation}
We employ this definition by choosing a region, evaluating $r$, determining the sign, and imposing this sign at the boundary point, i.e., on the region $[0,\theta]$ at the boundary point $\theta$ we obtain $\sgn(\gamma \theta) =1$, and on the region $[\theta,\infty)$ at the boundary point $\theta$ we obtain $\sgn(\gamma \theta -\delta)$.
We call this a \emph{wall labeling}.

The parameter graph, described below, allows us to construct all wall labelings possible for the network.
Observe that over all of the parameter space, there are two possible sign combinations, which correspond to parameter combinations that satisfy the inequalities
\begin{equation}
\label{ineq:1D_carrying_capacity}
0 < \gamma \theta < \delta  
\end{equation}
and
\begin{equation}
\label{ineq:1D_unbounded_growth}
0 < \delta < \gamma \theta,
\end{equation}
respectively (see Figure~\ref{fig:1D_network_signs}B). The inequalities  \eqref{ineq:1D_carrying_capacity} and \eqref{ineq:1D_unbounded_growth} provide the above-mentioned explicit decomposition of the parameter space, which we represent with the parameter graph. In particular, each sign combination is represented by one node; an edge is drawn between them to show that they share a co-dimension $1$ hypersurface (see Figure \ref{fig:1D_network_signs}B). The conclusion is that within our modeling framework, there are only two scenarios of dynamics that can occur.
If we assume that the parameters satisfy \eqref{ineq:1D_carrying_capacity}, then the population grows when it is below $\theta$ as it experiences no limitation, but declines when it is larger than $\theta$, since intraspecific competition is larger than growth.  For purposes of visualization, we use a right arrow to indicate a positive sign and a left arrow to indicate a negative sign. Thus, all dynamics point inward to the critical threshold $\theta$ (see arrows of Figure~\ref{fig:1D_network_signs}C; top). 
In contrast, assuming that the parameters satisfy \eqref{ineq:1D_unbounded_growth} leads to potentially unbounded growth (see Figure \ref{fig:1D_network_signs}C; bottom). For the sake of simplicity, we will assume the Malthusian viewpoint for prey populations and that they can not grow arbitrarily large, i.e. \eqref{ineq:1D_carrying_capacity} \citep{malthus1817essay}. In particular, in this paper, we will not analyze the dynamics that arise from \eqref{ineq:1D_unbounded_growth} because of its lack of ecological relevance, although $\wendy$ is capable of computing these as well. 
 
\begin{figure}[!htbp]
\centering
\begin{picture}(300,100)

\put(-60,0){{\bf \large A}}
\put(-55,10){
\begin{tikzpicture}
\draw[thick] (0,0)--(5,0);
\filldraw[black] (0,0) circle (2pt);
\filldraw[black] (1.67,0) circle (2pt);
\filldraw[black] (3.33,0) circle (2pt);
\filldraw[black] (5,0) circle (2pt);
\node at (.85,0.5) {$\kappa_0$};
\node at (2.5,0.5) {$\kappa_1$};
\node at (4.15,0.5) {$\kappa_2$};
\node[] (1) at (0,0) {};
\draw[->,thick] (1.3,0) -- (1.5,0);
\draw[->,thick] (1.78,0) -- (2.0,0);
\draw[->,thick] (3.23,0) -- (3.03,0);
\draw[->,thick] (3.73,0) -- (3.53,0);
\end{tikzpicture}
}

\put(135,0){{\bf \large B}}
\put(130,0){
\begin{tikzpicture}[thick, scale=0.9]
\node[ellipse, fill=white, draw] (2) at (0,2) {$\kappa_0$};
\node[ellipse, fill=white, draw] (3) at (2,2) {$\kappa_2$};
\node[ellipse, fill=white, draw] (4) at (1,0) {$\kappa_1$};
\draw[->, shorten <= 2pt, shorten >= 2pt] (2) to  (4);
\draw[->, shorten <= 2pt, shorten >= 2pt] (3) to  (4);
\draw[->, shorten <= 2pt, shorten >= 2pt] (2) to [out = 150, in = 210, looseness=5] (2); 
\draw[->, shorten <= 2pt, shorten >= 2pt] (4) to [out = 150, in = 210, looseness=5] (4); 
\end{tikzpicture}
}

\put(250,0){{\bf \large C}}
\put(270,8){
\begin{tikzpicture}[thick, scale=0.9]
\node[ellipse, fill=white, draw] (2) at (0,2) {$\kappa_0$};
\node[ellipse, fill=white, draw] (4) at (0,0) {$\kappa_1$};
\draw[->, shorten <= 2pt, shorten >= 2pt] (2) to  (4);
\end{tikzpicture}
}
\end{picture}
\caption{\footnotesize  {\bf Panel A}: the cell complex  and coarse dynamics of a population exhibiting carrying-capacity-like behavior. {\bf Panel B}: these dynamics are converted into a state transition graph; directed arrows from $\kappa_j \rightarrow \kappa_i$, where $i\neq j$, indicate a transition from cell $j$ to cell $i$; self-directed arrows from $\kappa_i \rightarrow \kappa_i$ indicate recurrent dynamics (i.e., a population can or will remain in the cell for all time). {\bf Panel C}: these dynamics are distilled into an $\sMG$, a simple graph encoding only the recurrent cells. Panels {\bf A} through {\bf C} thus outline how $\wendy$ analyzes a one-species community with intraspecific competition being larger than growth; the ultimate output is a set of $\sMG$s that distill and encode the long-term population dynamics. The bottom node of each $\sMG$ is where species' dynamics tend toward in the long-term.}
\label{fig:1D_Morse}
\end{figure}
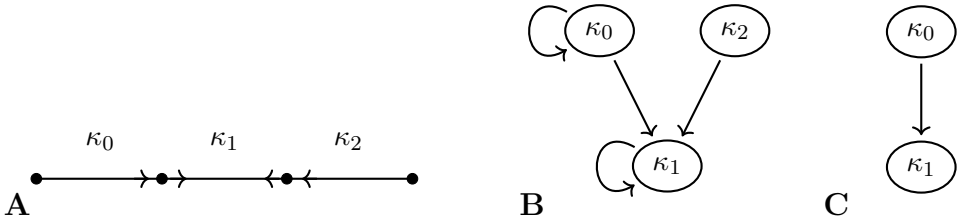

The second step in building the combinatorial model of the dynamics involves the construction of the desired cell complex (see Figure~\ref{fig:1D_Morse}A). In the context of parameters that satisfy \eqref{ineq:1D_carrying_capacity}, the cell complex consists of three cells (abstract intervals) $\kappa_0$, $\kappa_1$, and $\kappa_2$, where we identify $\kappa_0$ with the extinction set (i.e., the origin $x=0$) and small population densities below $\theta$ ($0 \leq x < \theta$), $\kappa_1$ with the critical threshold $\theta$, and $\kappa_2$ with the unbounded density region past the threshold $\theta$ ($x>\theta$). The population's tendency to grow or decline is indicated by the arrows in Figure~\ref{fig:1D_Morse}A. 

Using these wall labelings, we build a directed graph, the state transition graph (STG),  whose nodes are the cells (see Figure \ref{fig:1D_Morse}B): The link $\kappa_0 \to \kappa_0$ is justified because extinction is a fixed state and $\kappa_0$ is identified with the origin; $\kappa_0\to \kappa_1$, because small population levels increase without intraspecific competition; $\kappa_1\to \kappa_1$, because both wall labelings point into $\kappa_1$; and $\kappa_2 \to \kappa_1$, because large populations collapse when intraspecific competition overcomes growth. From the STG, we identify the Morse graph (MG) by identifying the recurrent nodes (i.e., the strongly connected components of the STG) $\kappa_0$ and $\kappa_1$; we then note that there is a path from $\kappa_0$ to $\kappa_1$ and denote this accordingly with a directed arrow from $\kappa_0 \rightarrow \kappa_1$ (see Figure \ref{fig:1D_Morse}C). Roughly, we see two ways this species' population will change over time: either carrying capacity eventually constrains its dynamics (i.e., enter cell $\kappa_1$ and stay permanently within this cell) or, if the population is initialized at $0$ density, it will forever remain at $0$ density (i.e., remain extinct). The importance of the MG is that it characterizes global long-term population dynamics. A node in a MG that is not the source of any directed arrow is called a \emph{minimal node} (see Figure \ref{fig:1D_Morse}C; the bottom node). Minimal nodes characterize attractors or sinks for the community dynamics, i.e., where long-term population dynamics will converge, and thus are of utmost ecological importance. 

\subsection{Two-species example: predator-prey model}
\label{sec:2species}

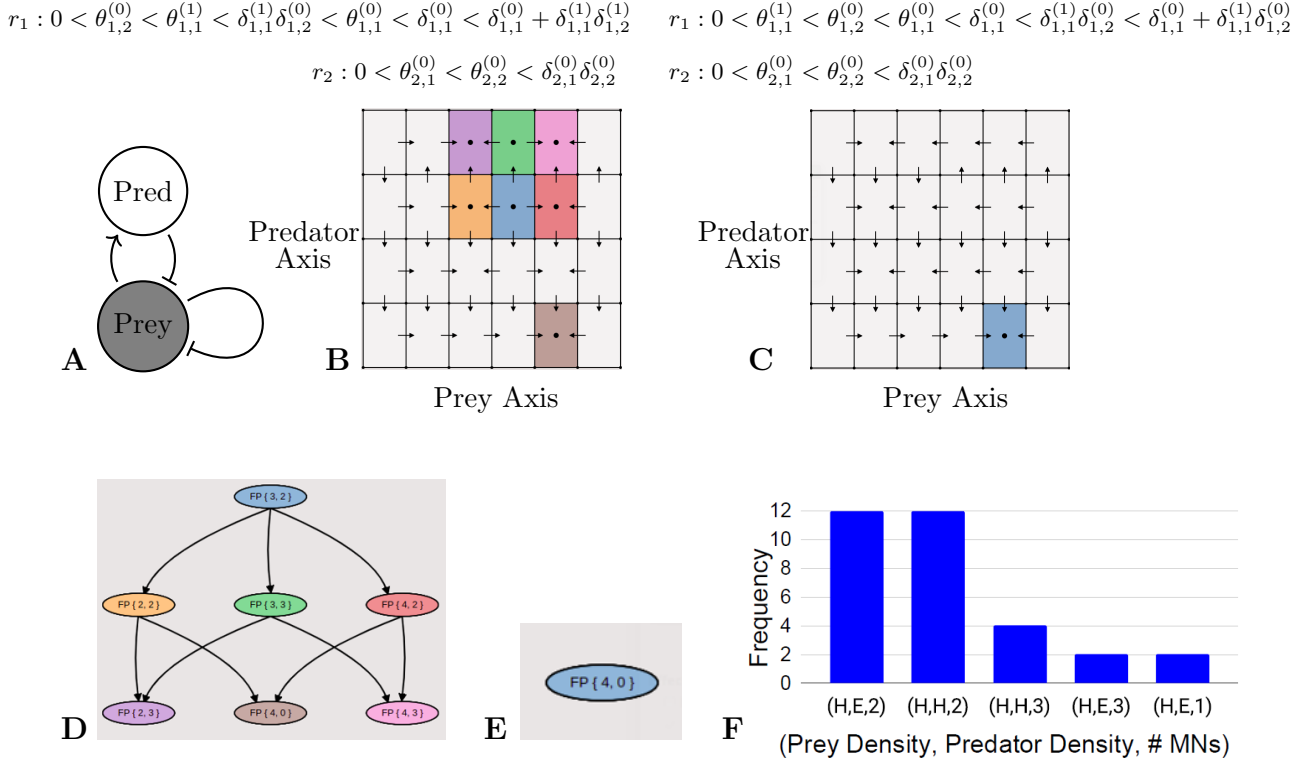
\begin{figure}[!htbp]
	\begin{picture}(350,315)
            

            \put(20,160){\bf \large A}
		\put(30,150){
			\begin{tikzpicture}[thick, scale=0.9]
				
				\node[circle, fill=gray, draw, minimum size=1cm] (1) at (0,0) {Prey};
				\node[circle, fill=white, draw, minimum size=1cm] (2) at (0,2) {Pred};

				\draw[-|, shorten <= 2pt, shorten >= 2pt] (1) to [out=30, in=-25, looseness=7] (1);
				\draw[->, shorten <= 2pt, shorten >= 2pt] (1) to [bend left](2);
				\draw[-|, shorten <= 2pt, shorten >= 2pt] (2) to [bend left](1);
				
		\end{tikzpicture}}

		\put(120,160){{\bf \large B}}
		\put(130,160){
			\includegraphics[scale = 0.25]{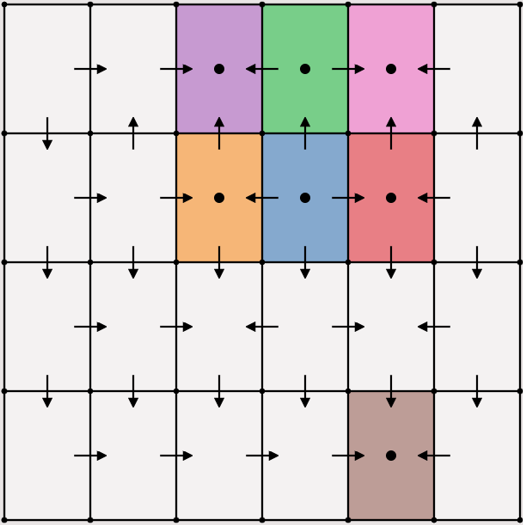}
		}

		\put(150,140){
			\begin{tikzpicture}[thick, scale = 0.9]
				\node at (0,0) {\textrm{ Prey Axis }};
			\end{tikzpicture}
		}
		
		\put(80,205){
			\begin{tikzpicture}[thick, scale = 0.9]
				\node at (0,0) {\textrm{ Predator }};
			\end{tikzpicture}
		}
		
		\put(90,195){
			\begin{tikzpicture}[thick, scale = 0.9]
				
				\node at (0,0) {\textrm{ Axis }};
				
			\end{tikzpicture}
		}

            \put(0,290){\footnotesize $r_1 : 0 < \theta_{1,2}^{(0)} < \theta_{1,1}^{(1)} < \delta_{1,1}^{(1)}\delta_{1,2}^{(0)} < \theta_{1,1}^{(0)} < \delta_{1,1}^{(0)} < \delta_{1,1}^{(0)} + \delta_{1,1}^{(1)} \delta_{1,2}^{(1)}$}
            
            \put(115,270){\footnotesize $r_2 : 0 <  \theta_{2,1}^{(0)} < \theta_{2,2}^{(0)} < \delta_{2,1}^{(0)} \delta_{2,2}^{(0)}$}
		
		\put(280,160){{\bf \large C}}
		\put(300,160){
			\includegraphics[scale = 0.25]{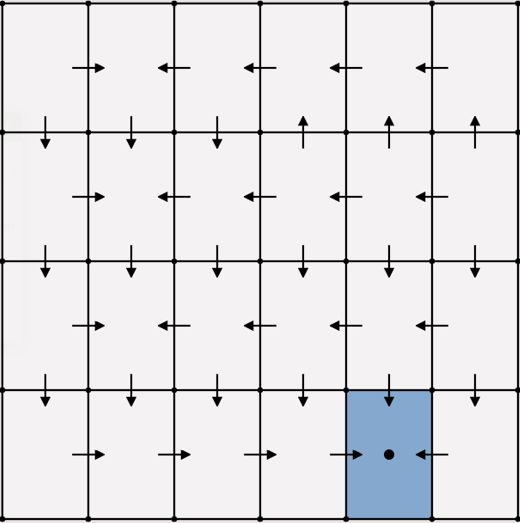}
		}
		
		\put(250,205){
			\begin{tikzpicture}[thick, scale = 0.9]
				
				\node at (0,0) {\textrm{ Predator }};
			\end{tikzpicture}
		}
		
		\put(260,195){
			\begin{tikzpicture}[thick, scale = 0.9]
				
				\node at (0,0) {\textrm{ Axis }};
				
			\end{tikzpicture}
		}
		
		\put(320,140){
			\begin{tikzpicture}[thick, scale = 0.9]
				
				\node at (0,0) {\textrm{ Prey Axis }};
				
			\end{tikzpicture}
		}

            \put(250,290){\footnotesize $r_1 : 0 < \theta_{1,1}^{(1)} < \theta_{1,2}^{(0)} < \theta_{1,1}^{(0)} < \delta_{1,1}^{(0)} < \delta_{1,1}^{(1)}\delta_{1,2}^{(0)}< \delta_{1,1}^{(0)} + \delta_{1,1}^{(1)}\delta_{1,2}^{(0)}$}
            \put(250,270){\footnotesize $r_2 : 0 < \theta_{2,1}^{(0)} < \theta_{2,2}^{(0)} < \delta_{2,1}^{(0)}\delta_{2,2}^{(0)}$}
		
		\put(20,20){{\bf \large D}}
		\put(30,20){
			\includegraphics[scale = 0.42]{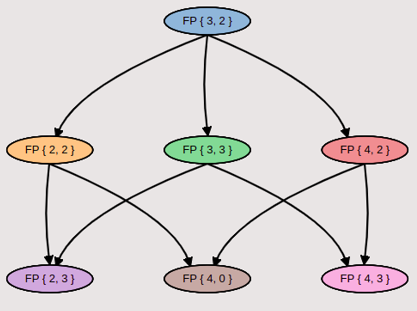}
		}
		
		\put(180,20){{\bf \large E}}
		\put(190,20){
			\includegraphics[scale = 0.65]{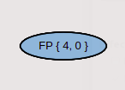}
		}
		
		\put(270,20){{\bf \large F}}
		\put(275,10){
			\includegraphics[scale = 0.37]{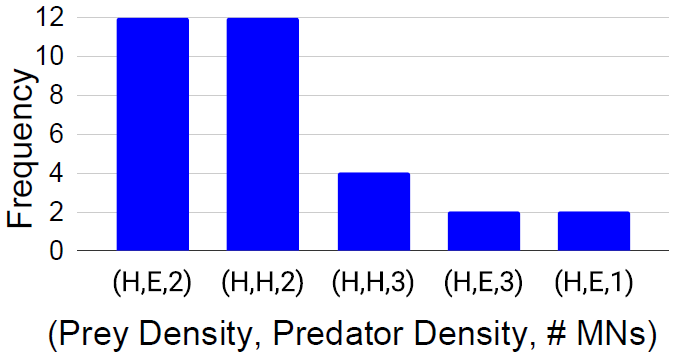}
		}
	\end{picture}
\caption{\footnotesize
{\bf Panel A}: Trophic network consisting of a prey (Node Prey) and predator (Node Pred). The grey indicates that in the absence of the predator population and intraspecific competition, the species increases in density with available background resources. Alternatively, the predator population declines when the prey population is sufficiently small and is colored white. The blunt self-edge and incoming edge on Node Prey indicate the prey declines from intraspecific competition and predation, respectively. The incoming pointed arrow to Node Pred indicates the predator benefits from the prey. {\bf Panels B and C}: two qualitatively different dynamics that can arise from different relative orderings of the parameters, the strength of predation, growth rates of the predator and prey, and strength of intraspecific competition of the prey.  Each set of inequalities -- corresponding to $1$ of the parameter nodes -- determines the signs of $r_1$ and $r_2$ and the direction of the dynamics (arrows) for the prey and the predator, respectively, on the boundaries of each cell. This collection of arrows forms a state transition graph (STG). {\bf Panels D and E}: corresponding Morse Graphs ($\sMG$s) of STGs in panels {\bf B} and {\bf C}.
Observe that for {\bf Panel D} the Morse graph has multiple minimal node. {\bf Panel F}: percent frequency of each $\minnode$ label that arises: $(\sH,\sE,2)$ and $(\sH,\sE,3)$ indicate the existence of at least one attracting cell in which the predator is extinct and the prey is present and high in density. $\minnode$ labels $(\sH, \sH,2)$ and $(\sH, \sH,3)$ indicate the existence of at least one local attracting cell where the predator and prey were both present and high in density. $\minnode$ label $(\sH,\sE,1)$ indicates the sole parameter node in which there was only one minimal node in the corresponding Morse graph; here, only the prey persists, regardless of the predator's starting density.}
\label{fig:2species}
\end{figure}


While the single-species example provides intuition concerning the implementation of $\wendy$, the power of our framework lies in its ability to fully analyze multispecies population networks. Before doing so for a real ecological network in the next section, let us consider now a theoretical two-species predator-prey community. For this example, the associated interaction network consists of two nodes: one for the prey and one for the predator (see Figure \ref{fig:2species}A). The arrow pointing from the prey node to the predator node indicates that the predator experiences a positive contribution to its growth, while the blunt edge from the predator node to the prey node indicates that the predator diminishes the prey population. As in the single-species example, the blunt self-edge at the prey node indicates intraspecies competition. 

We assume that the growth rate for the prey in the absence of any other species is $\gamma_1 > 0$, and that it experiences intraspecific competition at a rate of $\delta_{1,1}^{(0)} > 0$. Here, intraspecific competition only impacts the observable dynamics whenever the prey density surpasses a threshold $\theta_{1,1}^{(0)} > 0$. Additionally, the prey population’s growth rate is reduced by $\delta_{1,1}^{(1)} \delta_{1,2}^{(0)} > 0$ whenever the prey's population exceeds $\theta_{1,1}^{(1)} > 0$ \emph{and} the predator's population exceeds $\theta_{1,2}^{(0)} > 0$, and is otherwise unaffected. Regarding the predator, we assume a growth rate $-\gamma_2$  in the absence of any other species (with $\gamma_2 > 0$, i.e., population declines, indicated with white color for the node). If the density of the predator is below a threshold $\theta_{2,2}^{(0)} > 0$ \emph{or} the density of the prey is below a threshold $\theta_{2,1}^{(0)} > 0$, then we assume that no predation takes place. If both predator \emph{and} prey densities exceed the respective thresholds, then predation contributes to predator growth at a rate $\delta_{2,2}^{(0)} \delta_{2,1}^{(0)} > 0$. In other words, we assume that the prey population is not predated upon when rare in density and, in this case, that the predator population receives no benefit from the prey population and consequently declines. Even if the prey population is sufficiently large, a rare predator population is unable to successfully attack and consume the prey (e.g., the prey population can avoid or defend itself against a low number of predators). It is only if both populations are sufficiently large that the prey population is diminished by the predator and the predator population is bolstered by the prey. This translates mathematically into the following growth rate for the prey, $r_1$, and the predator's growth rate, $r_2$ (see Methods for details):

\begin{align}
r_1(x) &= \gamma_1 x_1 -
\begin{cases}
0, & \text{if}~ x_1 < \theta_{1,1}^{(0)} \\
\delta_{1,1}^{(0)}, & \text{if}~ x_1 > \theta_{1,1}^{(0)}
\end{cases}
- \left(
\begin{cases}
0, & \text{if}~ x_1 < \theta_{1,1}^{(1)} \\
\delta_{1,1}^{(1)}, & \text{if}~ x_1 > \theta_{1,1}^{(1)}
\end{cases}
\right)
\left(
\begin{cases}
0, & \text{if}~ x_2 < \theta_{1,2}^{(0)} \\
\delta_{1,2}^{(0)}, & \text{if}~ x_2 > \theta_{1,2}^{(0)}
\end{cases}
\right) \label{eqtn:preygrowthrate_in_predprey}
\\
r_2(x) &= - \gamma_2 x_2 +
\left(
\begin{cases}
0, & \text{if}~ x_1 < \theta_{2,1}^{(0)} \\
\delta_{2,1}^{(0)}, & \text{if}~ x_1 > \theta_{2,1}^{(0)}
\end{cases}
\right)
\left(
\begin{cases}
0, & \text{if}~ x_2 < \theta_{2,2}^{(0)} \\
\delta_{2,2}^{(0)}, & \text{if}~ x_2 > \theta_{2,2}^{(0)}
\end{cases}
\right).
\end{align}

Since there are two species, the finite cell complex $\cX$ over which we study the community's coarse population dynamics is $2$-dimensional ($k=2$), and thus each density region of $\cX$ is a rectangle. Similarly to the one-species case, the rate of change of the population density is calculated by choosing a rectangle, evaluating $r_1$ and $r_2$, determining the signs, and imposing the sign of $r_1$ and $r_2$ on the vertical and horizontal edges, respectively. This determines the wall labelings.

Implicit in the previous sentence and of fundamental importance is the requirement that the sign of the rate function is non-zero over the edges of the rectangles. This leads to sets of inequalities (analogous to inequality \eqref{ineq:1D_carrying_capacity}) that decompose parameter space $[0,\infty)^{17}$  into 96 parameter regions. From those regions, only $16$ meet our assumption that the prey reach carrying capacity in the absence of the predator (see two examples of such regions in Figure~\ref{fig:2species}B,C). Thus, for trophic networks with limited interactions like the focal example, $\wendy$ provides explicit precomputed decompositions of the parameter space. For more complicated cases, such as the real-world example discussed in the next Section~\ref{sec:coral}, $\wendy$ provides algorithms for sampling from the parameter graph.

To characterize the dynamics, we choose a single parameter region, identify the cell complex, and use the wall labelings to construct the STG. For purposes of visualizing the STG, we used right and up arrows to indicate positive growth and left and down arrows to indicate negative growth (see Figure~\ref{fig:2species}B and C). From the STG, we compute the associated MG to provide information about the global population dynamics. As shown in Figure~\ref{fig:2species}D and E, the result is a graph typically small compared to the STG, from which we can extract the most fundamental dynamics information, namely the minimal nodes that identify the long-term attractors of the predator-prey cell complex $\cX$. 

To this end, we use \emph{minimal node labels} (MNLs) that are defined by three indices as follows. The first column of cells of the finite cell complex $\cX$, i.e., rectangles parallel to the vertical predator axis (see e.g. Figures 3B and C), are declared to be \emph{extinction cells for the prey}; the adjacent column is identified as  \emph{low prey population}; and all other cells are considered to be \emph{high prey populations}. Thus, the first index of the MNL is denoted by $\sE$, $\sL$, or $\sH$ if the associated cell corresponds to extinction, low, or high prey population, respectively. The second index of the MNL is defined similarly but based on the rows of cells of the finite cell complex $\cX$ and captures the population levels of the predator. The last index of the MNL denotes how many minimal nodes there are in the MG: the index is $1$ if there is only one minimal node, $2$ if only two, and $3$ if there are three or more minimal nodes. Note that allowing for greater precision (e.g., explicitly labeling $4$, $5$, or more minimal nodes) can be done but also creates more MNLs that need to be tracked which can become cumbersome for larger-species communities.

To obtain results for our predator-prey example, we compute the MG for each parameter region and extract the associated MNLs. In total, we found 5 different MNLs: $(\sE,\sH,2)$ with occurrence 40\%, $(\sH,\sH,2)$ with $40\%$, $(\sE,\sH,3)$ with $6.7\%$, $(\sH,\sH,3)$ with 6.7\%, and $(\sE,\sH,1)$ with 6.7\%.
In particular, approximately $53\%$ of the MNLs indicate a locally attracting cell in which the prey is present but the predator is not. This is not to be misinterpreted as the predator going extinct in $53\%$ of cases. Rather, the predator goes extinct and the prey eventually reaches carrying capacity  if and only if the predator starts in a sufficiently low density region (see Figure \ref{fig:2species}B;D).
In fact, the MNLs indicate that the prey-only cell (i.e., the predator goes extinct eventually regardless of its initial density cell; see Figure \ref{fig:2species}C;E) is globally attracting in only $6.7\%$ of the cases. Similarly, in approximately $47\%$ of the MNLs there is at least one locally attracting cell in which the predator and prey are both present; however, long-term dynamics coarsely flow to these cells if and only if the predator starts in a sufficiently high density cell (e.g., $1$st row; $2$nd through $5$th columns of the cell complex in Figure \ref{fig:2species}B). In other words, the majority of these results exhibit bistability, i.e. there are alternative stable states for the community.

Note that, while there are $16$ nodes in the parameter graph, there are only 5 distinct MNLs.
This can be viewed as an advantage of our representation of dynamics: whereas traditional perspectives based on invariant sets require precise detailed understanding of nonrobust dynamic structures (see \cite{ugarcovici:weiss}), our combinatorial methods identify robust structures (presumably more easily observable experimentally) that persist over large regions of parameter space \citep{database,bush:gameiro:harker:kokubu:mischaikow:obayashi:pilarczyk,bush:mischaikow}.

\section{Indonesian coral reef mutualistic network}
\label{sec:coral}

To demonstrate the applicability of $\wendy$ and motivated by \citep{ricciardi2010assemblage} we consider an existing mutualistic community of $7$ anemonefish species (\emph{Amphiprion clarkii}, \emph{A. melanopus}, \emph{A. ocellaris}, \emph{A. perideraion}, \emph{A. polymnus}, \emph{A. sandaracinos}, \emph{Premnas biaculeatus}) and $8$ anemone species (\emph{Entacmaea quadricolor}; \emph{Macrodactyla doreensis} (Fam. Actiniidae); \emph{Heteractis aurora}; \emph{H. crispa}; \emph{H. magnifica}; \emph{H. malu}; \emph{S. haddoni}; \emph{S. mertensii} (Fam. Stichodactylidae)) in Indonesia. We aim to understand the long-term dynamics of the community. To this end, we use $\wendy$ in combination with the experimental data reported in \cite{ricciardi2010assemblage}, narrowing down framework choices attending to biological relevance and assumptions and computational constraints.
Thus, we first ask: What interaction networks are most appropriate for the analysis of potential dynamics?

Similar to the previous example, we assume that the net growth rates for the anemone in isolation are positive and thus represented by grey nodes, and those for anemonefish negative (white nodes).
We assume that there is no interaction (i.e., no edge) between anemones, but that each anemone is subject to intraspecific competition (i.e., blunt self-edge). Based on the discussion in \citep{ricciardi2010assemblage}, we assume that any interaction between anemonefish is competitive (blunt edges) and any interaction between anemonefish and anemone is mutualistic (pointed edges).

Mathematically, we model the associated rate functions as follows (for details see the Appendix).
If there is a pointed edge from anemonefish species $i$ to anemone species $j$, then we assume that the anemonefish growth rate is increased from the interaction by the rate
$\delta_{i,i}^{(m)} \delta_{j,i}^{(0)}$
(where $m$ denotes the $m$-th interaction species $i$ has with another species) but only if the anemonefish has density greater than $\theta_{i,i}^{(m)}$ and the anemone species has greater density than $\theta_{j,i}^{(0)}$.
Because we are assuming a mutualistic interaction, a pointed edge from anemonefish species $i$ to anemone species $j$ implies the existence of a pointed edge from anemone species $j$ to anemonefish species $i$. Similarly, the anemone growth rate is increased from the interaction by the rate $\delta_{j,j}^{(\widetilde m)} \delta_{i,j}^{(0)}$
but only if the anemone has density greater than $\theta_{j,j}^{(\widetilde m)}$ and the anemonefish has density greater than $\theta_{i,j}^{(0)}$.
If anemonefish species $i$ and $j$ are competing (blunt edges), then we assume species $i$ has an impact on species $j$ at a rate $-\delta_{j,j}^{(m)}\delta_{j,i}^{(0)}$, but only if their densities are higher than $\theta_{j,j}^{(m)}$ and $\theta_{j,i}^{(0)}$, respectively. 

Although occasional encounters between anemone and anemonefish may be due to just chance or large populations, a reasonable assumption is that \emph{ceteris paribus}, the greater the number of visits the more likely an interaction is.
With this in mind, for our analyses, we consider a family of interaction networks by including edges between an anemonefish to an anemone whenever the frequency of visits as reported in \cite[Table 2]{ricciardi2010assemblage} and per-species preference \cite[Table 3]{ricciardi2010assemblage} exceeds those thresholds set based on computational constraints (see below). Competitive edges between the anemonefish are determined by yet another set of thresholds \cite[Table 4]{ricciardi2010assemblage}.

\begin{figure}[!htbp]
	\begin{picture}(400,470)
	
		\put(0,300){ 
		\begin{tikzpicture}[thick, scale=0.75]
		
		\node[circle, fill=white, draw, minimum size=1cm] (1) at (0,3.5) {\emph{A.cl.}};
		\node[circle, fill=white, draw, minimum size=1cm] (2) at (2,3.5) {\emph{A.pe.}};
		\node[circle, fill=white, draw, minimum size=1cm] (5) at (4,3.5) {\emph{A.oc.}};
		\node[circle, fill=white, draw, minimum size=1cm] (6) at (6,3.5) {\emph{A.sa.}};
		
		\node[circle, fill=gray, draw, minimum size=1cm] (8) at (0,0) {\emph{H.cr.}};
		\node[circle, fill=gray, draw, minimum size=1cm] (9) at (2,0) {\emph{H.ma.}};
		\node[circle, fill=gray, draw, minimum size=1cm] (10) at (4,0) {\emph{H.me.}};
		\node[circle, fill=gray, draw, minimum size=1cm] (12) at (6,0) {\emph{H.au.}};
		\node (0) at (6,1.5) {\bf \large N36};
		
		\draw[-|, shorten <= 2pt, shorten >= 2pt] (1) to [out=15, in=165, looseness=1] (2);
		\draw[-|, shorten <= 2pt, shorten >= 2pt] (2) to [out=185, in=-5, looseness=1] (1);
		\draw[-|, shorten <= 2pt, shorten >= 2pt] (1) to [out=30, in=150, looseness=1] (6);
		\draw[-|, shorten <= 2pt, shorten >= 2pt] (6) to [out=130, in=50, looseness=1] (1);
		
		\draw[->, shorten <= 2pt, shorten >= 2pt] (1) to [out=180, in=180, looseness=1] (8);
		\draw[->, shorten <= 2pt, shorten >= 2pt] (8) to [out=160, in=200, looseness=1] (1);
		\draw[->, shorten <= 2pt, shorten >= 2pt] (1) to [out=240, in=150, looseness=1] (10);
		\draw[->, shorten <= 2pt, shorten >= 2pt] (10) to [out=130, in=260, looseness=1] (1);
		\draw[->, shorten <= 2pt, shorten >= 2pt] (1) to [out=325, in=150, looseness=1] (12);
		\draw[->, shorten <= 2pt, shorten >= 2pt] (12) to [out=140, in=335, looseness=1] (1);
		
		\draw[->, shorten <= 2pt, shorten >= 2pt] (2) to [out=220, in=100, looseness=1] (8);
		\draw[->, shorten <= 2pt, shorten >= 2pt] (8) to [out=85, in=235, looseness=1] (2);
		\draw[->, shorten <= 2pt, shorten >= 2pt] (2) to [out=255, in=110, looseness=1] (9);
		\draw[->, shorten <= 2pt, shorten >= 2pt] (9) to [out=95, in=270, looseness=1] (2);
		
		\draw[->, shorten <= 2pt, shorten >= 2pt] (5) to [out=240, in=75, looseness=1] (9);
		\draw[->, shorten <= 2pt, shorten >= 2pt] (9) to [out=60, in=255, looseness=1] (5);
		
		\draw[->, shorten <= 2pt, shorten >= 2pt] (6) to [out=240, in=75, looseness=1] (10);
		\draw[->, shorten <= 2pt, shorten >= 2pt] (10) to [out=60, in=255, looseness=1] (6);
		
		\draw[-|, shorten <= 2pt, shorten >= 2pt] (8) to [out=290, in=250, looseness=7] (8);
		\draw[-|, shorten <= 2pt, shorten >= 2pt] (9) to [out=290, in=250, looseness=7] (9);
		\draw[-|, shorten <= 2pt, shorten >= 2pt] (10) to [out=290, in=250, looseness=7] (10);
		\draw[-|, shorten <= 2pt, shorten >= 2pt] (12) to [out=290, in=250, looseness=7] (12);
		
		\end{tikzpicture}}
	
	\put(250,300){ 
		\begin{tikzpicture}[thick, scale=0.75]
			
		\node[circle, fill=white, draw, minimum size=1cm] (1) at (0,3.5) {\emph{A.cl.}};
		\node[circle, fill=white, draw, minimum size=1cm] (2) at (2,3.5) {\emph{A.pe.}};
		\node[circle, fill=white, draw, minimum size=1cm] (5) at (4,3.5) {\emph{A.oc.}};
		\node[circle, fill=white, draw, minimum size=1cm] (6) at (6,3.5) {\emph{A.sa.}};
		
		\node[circle, fill=gray, draw, minimum size=1cm] (8) at (0,0) {\emph{H.cr.}};
		\node[circle, fill=gray, draw, minimum size=1cm] (9) at (2,0) {\emph{H.ma.}};
		\node[circle, fill=gray, draw, minimum size=1cm] (10) at (4,0) {\emph{H.me.}};
		\node[circle, fill=gray, draw, minimum size=1cm] (12) at (6,0) {\emph{H.au.}};
			\node (0) at (6,1.5) {\bf \large N30};
			
			\draw[-|, shorten <= 2pt, shorten >= 2pt] (1) to [out=30, in=150, looseness=1] (6);
			\draw[-|, shorten <= 2pt, shorten >= 2pt] (6) to [out=130, in=50, looseness=1] (1);
			
			\draw[->, shorten <= 2pt, shorten >= 2pt] (1) to [out=180, in=180, looseness=1] (8);
			\draw[->, shorten <= 2pt, shorten >= 2pt] (8) to [out=160, in=200, looseness=1] (1);
			\draw[->, shorten <= 2pt, shorten >= 2pt] (1) to [out=240, in=150, looseness=1] (10);
			\draw[->, shorten <= 2pt, shorten >= 2pt] (10) to [out=130, in=260, looseness=1] (1);
			\draw[->, shorten <= 2pt, shorten >= 2pt] (1) to [out=325, in=150, looseness=1] (12);
			\draw[->, shorten <= 2pt, shorten >= 2pt] (12) to [out=140, in=335, looseness=1] (1);
			
			\draw[->, shorten <= 2pt, shorten >= 2pt] (2) to [out=220, in=100, looseness=1] (8);
			\draw[->, shorten <= 2pt, shorten >= 2pt] (8) to [out=85, in=235, looseness=1] (2);
			\draw[->, shorten <= 2pt, shorten >= 2pt] (2) to [out=255, in=110, looseness=1] (9);
			\draw[->, shorten <= 2pt, shorten >= 2pt] (9) to [out=95, in=270, looseness=1] (2);
			
			\draw[->, shorten <= 2pt, shorten >= 2pt] (5) to [out=240, in=75, looseness=1] (9);
			\draw[->, shorten <= 2pt, shorten >= 2pt] (9) to [out=60, in=255, looseness=1] (5);
			
			\draw[->, shorten <= 2pt, shorten >= 2pt] (6) to [out=240, in=75, looseness=1] (10);
			\draw[->, shorten <= 2pt, shorten >= 2pt] (10) to [out=60, in=255, looseness=1] (6);
			
			\draw[-|, shorten <= 2pt, shorten >= 2pt] (8) to [out=290, in=250, looseness=7] (8);
			\draw[-|, shorten <= 2pt, shorten >= 2pt] (9) to [out=290, in=250, looseness=7] (9);
			\draw[-|, shorten <= 2pt, shorten >= 2pt] (10) to [out=290, in=250, looseness=7] (10);
			\draw[-|, shorten <= 2pt, shorten >= 2pt] (12) to [out=290, in=250, looseness=7] (12);
			
	\end{tikzpicture}}



\put(20,140){
	\begin{tikzpicture}[thick, scale=0.75]
		
		\node[circle, line width=1mm, draw = purple, fill=white, draw, minimum size=1cm] (1) at (0,3.5) {\emph{A.cl.}};
		\node[circle, fill=white, draw, minimum size=1cm] (2) at (2,3.5) {\emph{A.pe.}};
		\node[circle, fill=white, draw, minimum size=1cm] (5) at (4,3.5) {\emph{A.oc.}};
		\node[circle, line width=1mm, draw = purple, fill=white, draw, minimum size=1cm] (6) at (6,3.5) {\emph{A.sa.}};
		
		\node[circle, fill=gray, draw, minimum size=1cm] (8) at (0,0) {\emph{H.cr.}};
		\node[circle, fill=gray, draw, minimum size=1cm] (9) at (2,0) {\emph{H.ma.}};
		\node[circle, line width=1mm, draw = purple, fill=gray, draw, minimum size=1cm] (10) at (4,0) {\emph{H.me.}};
		\node[circle, line width=1mm, draw = purple, fill=gray, draw, minimum size=1cm] (12) at (6,0) {\emph{H.au.}};
		\node (0) at (6,1.5) {\bf \large N24};
		
		\draw[-|, draw = purple, shorten <= 2pt, shorten >= 2pt] (1) to [out=30, in=150, looseness=1] (6);
		\draw[-|, draw = purple, shorten <= 2pt, shorten >= 2pt] (6) to [out=130, in=50, looseness=1] (1);
		
		\draw[->, draw = purple, shorten <= 2pt, shorten >= 2pt] (1) to [out=240, in=150, looseness=1] (10);
		\draw[->, draw = purple, shorten <= 2pt, shorten >= 2pt] (10) to [out=130, in=260, looseness=1] (1);
		\draw[->, draw = purple, shorten <= 2pt, shorten >= 2pt] (1) to [out=325, in=150, looseness=1] (12);
		\draw[->, draw = purple, shorten <= 2pt, shorten >= 2pt] (12) to [out=140, in=335, looseness=1] (1);
		
		\draw[->, shorten <= 2pt, shorten >= 2pt] (2) to [out=220, in=100, looseness=1] (8);
		\draw[->, shorten <= 2pt, shorten >= 2pt] (8) to [out=85, in=235, looseness=1] (2);
		\draw[->, shorten <= 2pt, shorten >= 2pt] (2) to [out=255, in=110, looseness=1] (9);
		\draw[->, shorten <= 2pt, shorten >= 2pt] (9) to [out=95, in=270, looseness=1] (2);
		
		\draw[->, shorten <= 2pt, shorten >= 2pt] (5) to [out=240, in=75, looseness=1] (9);
		\draw[->, shorten <= 2pt, shorten >= 2pt] (9) to [out=60, in=255, looseness=1] (5);
		
		\draw[->, shorten <= 2pt, shorten >= 2pt, draw = purple] (6) to [out=240, in=75, looseness=1] (10);
		\draw[->, shorten <= 2pt, shorten >= 2pt, draw = purple] (10) to [out=60, in=255, looseness=1] (6);
		
		\draw[-|, shorten <= 2pt, shorten >= 2pt] (8) to [out=290, in=250, looseness=7] (8);
		\draw[-|, shorten <= 2pt, shorten >= 2pt] (9) to [out=290, in=250, looseness=7] (9);
		\draw[-|, draw = purple, shorten <= 2pt, shorten >= 2pt] (10) to [out=290, in=250, looseness=7] (10);
		\draw[-|, draw = purple, shorten <= 2pt, shorten >= 2pt] (12) to [out=290, in=250, looseness=7] (12);
		
\end{tikzpicture}}


\put(250,150){ 
	\begin{tikzpicture}[thick, scale=0.75]
		
		\node[circle, fill=white, draw, minimum size=1cm] (1) at (0,3.5) {\emph{A.cl.}};
		\node[circle, fill=white, draw, minimum size=1cm] (2) at (2,3.5) {\emph{A.pe.}};
		\node[circle, fill=white, draw, minimum size=1cm] (5) at (4,3.5) {\emph{A.oc.}};
		\node[circle, fill=white, draw, minimum size=1cm] (6) at (6,3.5) {\emph{A.sa.}};
		
		\node[circle, fill=gray, draw, minimum size=1cm] (8) at (0,0) {\emph{H.cr.}};
		\node[circle, fill=gray, draw, minimum size=1cm] (9) at (2,0) {\emph{H.ma.}};
		\node[circle, fill=gray, draw, minimum size=1cm] (10) at (4,0) {\emph{H.me.}};
		\node[circle, fill=gray, draw, minimum size=1cm] (12) at (6,0) {\emph{H.au.}};
		\node (0) at (6,1.5) {\bf \large N29};
		
		
		\draw[->, shorten <= 2pt, shorten >= 2pt] (1) to [out=180, in=180, looseness=1] (8);
		\draw[->, shorten <= 2pt, shorten >= 2pt] (8) to [out=160, in=200, looseness=1] (1);
		\draw[->, shorten <= 2pt, shorten >= 2pt] (1) to [out=240, in=150, looseness=1] (10);
		\draw[->, shorten <= 2pt, shorten >= 2pt] (10) to [out=130, in=260, looseness=1] (1);
		\draw[->, shorten <= 2pt, shorten >= 2pt] (1) to [out=325, in=150, looseness=1] (12);
		\draw[->, shorten <= 2pt, shorten >= 2pt] (12) to [out=140, in=335, looseness=1] (1);
		
		\draw[->, shorten <= 2pt, shorten >= 2pt] (2) to [out=220, in=100, looseness=1] (8);
		\draw[->, shorten <= 2pt, shorten >= 2pt] (8) to [out=85, in=235, looseness=1] (2);
		\draw[->, shorten <= 2pt, shorten >= 2pt] (2) to [out=255, in=110, looseness=1] (9);
		\draw[->, shorten <= 2pt, shorten >= 2pt] (9) to [out=95, in=270, looseness=1] (2);
		
		\draw[->, shorten <= 2pt, shorten >= 2pt] (5) to [out=240, in=75, looseness=1] (9);
		\draw[->, shorten <= 2pt, shorten >= 2pt] (9) to [out=60, in=255, looseness=1] (5);
		
		\draw[->, shorten <= 2pt, shorten >= 2pt] (6) to [out=240, in=75, looseness=1] (10);
		\draw[->, shorten <= 2pt, shorten >= 2pt] (10) to [out=60, in=255, looseness=1] (6);
		
		\draw[-|, shorten <= 2pt, shorten >= 2pt] (8) to [out=290, in=250, looseness=7] (8);
		\draw[-|, shorten <= 2pt, shorten >= 2pt] (9) to [out=290, in=250, looseness=7] (9);
		\draw[-|, shorten <= 2pt, shorten >= 2pt] (10) to [out=290, in=250, looseness=7] (10);
		\draw[-|, shorten <= 2pt, shorten >= 2pt] (12) to [out=290, in=250, looseness=7] (12);
		
\end{tikzpicture}}



\put(20,0){ 
	\begin{tikzpicture}[thick, scale=0.75]
		
		\node[circle, line width=1mm, draw = purple, fill=white, draw, minimum size=1cm] (1) at (0,3.5) {\emph{A.cl.}};
		\node[circle, fill=white, draw, minimum size=1cm] (2) at (2,3.5) {\emph{A.pe.}};
		\node[circle, fill=white, draw, minimum size=1cm] (5) at (4,3.5) {\emph{A.oc.}};
		\node[circle, line width=1mm, draw = purple, fill=white, draw, minimum size=1cm] (6) at (6,3.5) {\emph{A.sa.}};
		
		\node[circle, fill=gray, draw, minimum size=1cm] (8) at (0,0) {\emph{H.cr.}};
		\node[circle, fill=gray, draw, minimum size=1cm] (9) at (2,0) {\emph{H.ma.}};
		\node[circle, line width=1mm, draw = purple, fill=gray, draw, minimum size=1cm] (10) at (4,0) {\emph{H.me.}};
		\node[circle, line width=1mm, draw = purple, fill=gray, draw, minimum size=1cm] (12) at (6,0) {\emph{H.au.}};
		\node (0) at (6,1.5) {\bf \large N28};
		
		
		\draw[->, draw = purple, shorten <= 2pt, shorten >= 2pt] (1) to [out=240, in=150, looseness=1] (10);
		\draw[->, draw = purple, shorten <= 2pt, shorten >= 2pt] (10) to [out=130, in=260, looseness=1] (1);
		\draw[->, draw = purple, shorten <= 2pt, shorten >= 2pt] (1) to [out=325, in=150, looseness=1] (12);
		\draw[->, draw = purple, shorten <= 2pt, shorten >= 2pt] (12) to [out=140, in=335, looseness=1] (1);
		
		\draw[->, shorten <= 2pt, shorten >= 2pt] (2) to [out=220, in=100, looseness=1] (8);
		\draw[->, shorten <= 2pt, shorten >= 2pt] (8) to [out=85, in=235, looseness=1] (2);
		\draw[->, shorten <= 2pt, shorten >= 2pt] (2) to [out=255, in=110, looseness=1] (9);
		\draw[->, shorten <= 2pt, shorten >= 2pt] (9) to [out=95, in=270, looseness=1] (2);
		
		\draw[->, shorten <= 2pt, shorten >= 2pt] (5) to [out=240, in=75, looseness=1] (9);
		\draw[->, shorten <= 2pt, shorten >= 2pt] (9) to [out=60, in=255, looseness=1] (5);
		
		\draw[->, draw = purple, shorten <= 2pt, shorten >= 2pt] (6) to [out=240, in=75, looseness=1] (10);
		\draw[->, draw = purple, shorten <= 2pt, shorten >= 2pt] (10) to [out=60, in=255, looseness=1] (6);
		
		\draw[-|, shorten <= 2pt, shorten >= 2pt] (8) to [out=290, in=250, looseness=7] (8);
		\draw[-|, shorten <= 2pt, shorten >= 2pt] (9) to [out=290, in=250, looseness=7] (9);
		\draw[-|, draw = purple, shorten <= 2pt, shorten >= 2pt] (10) to [out=290, in=250, looseness=7] (10);
		\draw[-|, draw = purple, shorten <= 2pt, shorten >= 2pt] (12) to [out=290, in=250, looseness=7] (12);
		
\end{tikzpicture}}


\put(280,0){ 
	\begin{tikzpicture}[thick, scale=0.75]
		
		\node[circle, line width=1mm, draw=purple, fill=white, draw, minimum size=1cm] (1) at (0,3.5) {\emph{A.cl.}};
		\node[circle, fill=white, draw, minimum size=1cm] (2) at (2,3.5) {\emph{A.pe.}};
		\node[circle, fill=white, draw, minimum size=1cm] (5) at (4,3.5) {\emph{A.oc.}};
		\node[circle, line width=1mm, draw = blue, fill=white, draw, minimum size=1cm] (6) at (6,3.5) {\emph{A.sa.}};
		
		\node[circle, fill=gray, draw, minimum size=1cm] (8) at (0,0) {\emph{H.cr.}};
		\node[circle, fill=gray, draw, minimum size=1cm] (9) at (2,0) {\emph{H.ma.}};
		\node[circle, line width=1mm, draw = blue, fill=gray, draw, minimum size=1cm] (10) at (4,0) {\emph{H.me.}};
		\node[circle, line width=1mm, draw = purple, fill=gray, draw, minimum size=1cm] (12) at (6,0) {\emph{H.au.}};
		
		
		\draw[->, draw = purple, shorten <= 2pt, shorten >= 2pt] (1) to [out=325, in=150, looseness=1] (12);
		\draw[->, draw = purple, shorten <= 2pt, shorten >= 2pt] (12) to [out=140, in=335, looseness=1] (1);
		
		\draw[->, shorten <= 2pt, shorten >= 2pt] (2) to [out=220, in=100, looseness=1] (8);
		\draw[->, shorten <= 2pt, shorten >= 2pt] (8) to [out=85, in=235, looseness=1] (2);
		\draw[->, shorten <= 2pt, shorten >= 2pt] (2) to [out=255, in=110, looseness=1] (9);
		\draw[->, shorten <= 2pt, shorten >= 2pt] (9) to [out=95, in=270, looseness=1] (2);
		
		\draw[->, shorten <= 2pt, shorten >= 2pt] (5) to [out=240, in=75, looseness=1] (9);
		\draw[->, shorten <= 2pt, shorten >= 2pt] (9) to [out=60, in=255, looseness=1] (5);
		
		\draw[->, draw = blue, shorten <= 2pt, shorten >= 2pt] (6) to [out=240, in=75, looseness=1] (10);
		\draw[->, draw = blue, shorten <= 2pt, shorten >= 2pt] (10) to [out=60, in=255, looseness=1] (6);
		
		\draw[-|, shorten <= 2pt, shorten >= 2pt] (8) to [out=290, in=250, looseness=7] (8);
		\draw[-|, shorten <= 2pt, shorten >= 2pt] (9) to [out=290, in=250, looseness=7] (9);
		\draw[-|, draw = blue, shorten <= 2pt, shorten >= 2pt] (10) to [out=290, in=250, looseness=7] (10);
		\draw[-|, draw = purple, shorten <= 2pt, shorten >= 2pt] (12) to [out=290, in=250, looseness=7] (12);
		\node (0) at (6,1.5) {\bf \large N21};
		
\end{tikzpicture}}
		
\end{picture}
\caption{\footnotesize All networks that are analyzed in the Indonesian coral reef example. {\bf N36} is the full community with all thresholds of frequency of visits, preferences, and niche overlap values set to $0$. {\bf N30}, {\bf N24}, {\bf N29}, {\bf N28}, and {\bf N21} are all uniquely different subconfigurations of {\bf N36} of interactions between anemone and anemonefish species. The top row of each network (left to right) consists solely of anemonefish species \emph{A. Clarkii}, \emph{A. perideraion}, \emph{A. ocellaris}, and \emph{A. sandaracinos}, respectively. The bottom row of each network (left to right) consists solely of sea anemone species \emph{H. crispa}, \emph{H. magnifica}, \emph{S. mertensii}, and \emph{H. aurora}, respectively. Each sea anemone species is assumed to increase in density in the absence of other species and intraspecific competition and are thus shaded grey. Edges have been colored purple (in {\bf N21}, {\bf N24}, and {\bf N28}) and blue (in {\bf N21}) to denote isolated subcommunities within the configuration (i.e., each isolated subcommunity does not interact with those of a different color).} 
\label{fig:poset_full_sub_communities}
\end{figure}
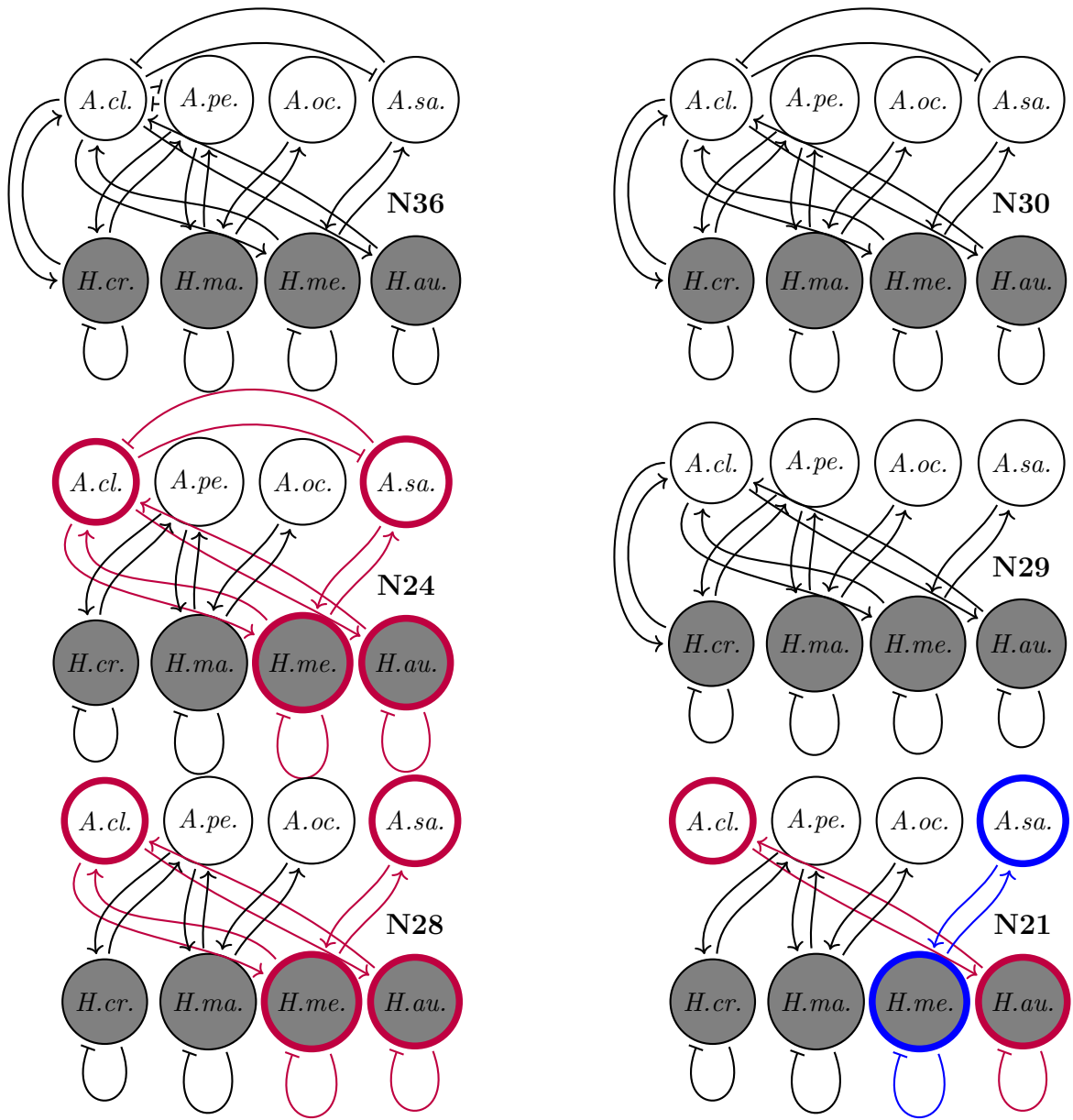

Sweeping over all possible thresholds gives rise to $61$ distinct interaction networks. The minimal network, obtained by setting each threshold beyond the values in the above-mentioned tables has $15$ nodes but no edges between nodes. The maximal network, obtained by setting each threshold equal to zero, also has $15$ nodes, but now each anemonefish competes with every other anemonefish and every anemonefish has a mutualistic interaction with each anemone. The minimal network is of no biological interest, and the maximal network (also of limited biological interest) exceeds our computational capabilities (see below).
Thus, to find intermediate networks, we impose two criteria. First, every species has to interact with at least one other species; isolated species have no direct effect on the community and vice versa. Second, a bound on the computation time is needed, to compute a sufficient number of MG that provides confidence in the parameter-wide description of asymptotic dynamics. 
These two criteria, given our computational resources (a cluster with $100$ nodes),  result in $6$ interaction networks: ${\bf N36}$ (consisting of $4$ anemonefish and $4$ sea anemone species, i.e., $k=8$) and its subnetworks (see Figure~\ref{fig:poset_full_sub_communities}).

In particular, {\bf N30} is a subconfiguration of {\bf N36}; {\bf N24} 
and {\bf N29} are subconfigurations of {\bf N30} (but not of each other); {\bf N28} is a subconfiguration of both {\bf N24} and {\bf N29}; finally, {\bf N21} is a subconfiguration of {\bf N28}. Subnetworks {\bf N29}, {\bf N30}, and {\bf N36} form one $8$-species community. 
{\bf N24} and {\bf N28} are subnetwork configurations composed of two isolated subcommunities, each consisting of $2$ anemonefish species and $2$ anemone species. 
The most disconnected subnetwork, {\bf N21}, is composed of three isolated subcommunities: two $2$-species subcommunities composed of an anemonefish species and an anemone species, and a $4$-species subcommunity of two anemonefish species and two anemone species. 

Computational constraints prevent us from lowering any of the thresholds (the computations took about $8.5$ hours for each network), as this leads to larger networks with significantly more edges and, consequently, dramatically increases the number of parameter regions and also the cost of computing the MGs.
Ecologically, we are then limited to analyzing a community that has, at most, a moderate number of interactions for any given species, and therefore our analyses do not precisely capture the impact that generalists such as \emph{Amphiprion clarkii} can have on the coral reef community. Nevertheless, $\wendy$ provides a ``broad strokes'' understanding of how mutualistic and antagonistic interactions, and the lack thereof, impact the long-term dynamics of this marine system. 

Analogous to the predator-prey example, from the MGs for each interaction network we extract MNLs of the form $(S_1, S_2, S_3, S_4, S_5, S_6, S_7, S_8, n)$,
where $S_1$, $S_2$, $S_3$,  and $S_4$ are anemonefish species \emph{A. Clarkii}, \emph{A. perideraion}, \emph{A. ocellaris}, and \emph{A. sandaracinos}, respectively. $S_5$, $S_6$, $S_7$,  and $S_8$ are anemone species \emph{H. crispa}, \emph{H. magnifica}, \emph{S. mertensii}, and \emph{H. aurora}, respectively. 
Similar to the two-species scenario, $S_i \in \{\sE, \sL, \sH \}$, for $i \in \{1,2,\ldots,8\}$, denotes whether an MNL corresponds to a cell in which species $i$ is extinct,  has a low presence, or has a high presence. Like in the previous example, the final entry $n$ equals $1$ if the MG only has one minimal node, $2$ if it only has two minimal nodes, or $3$ if it has three or more minimal nodes.

Finally, we tracked whether or not oscillations occurred, that is, whenever the $\sMG$s contained minimal nodes that represented dynamics in which at least one species recurrently visited separate density regions.
Note that we excluded these instances, as they were rare.

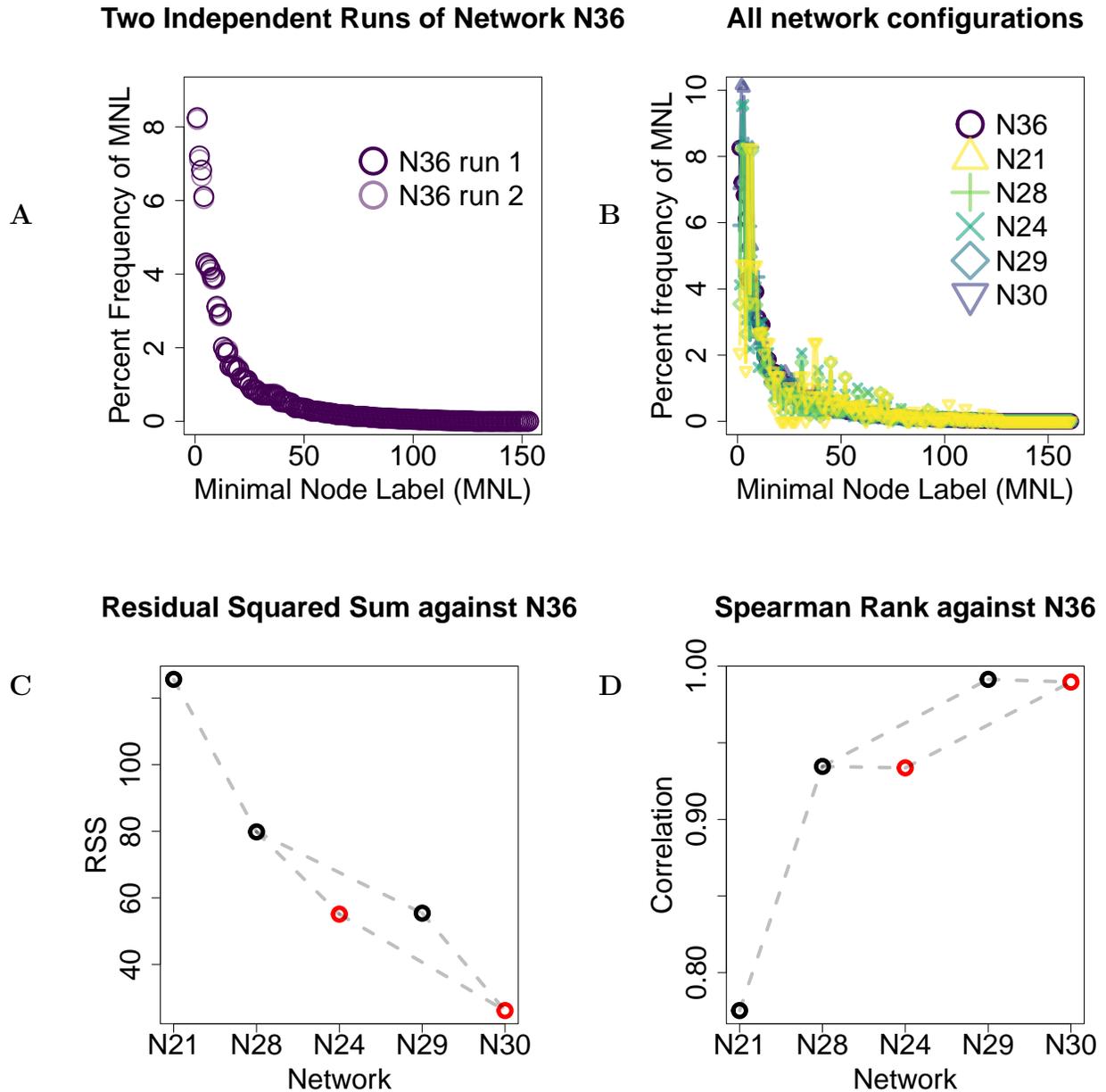
\begin{figure}[!htbp]
\begin{picture}(400,500)
\put(0,400){{\bf \large A}}
\put(20,260){ 
\includegraphics[scale = 0.5]{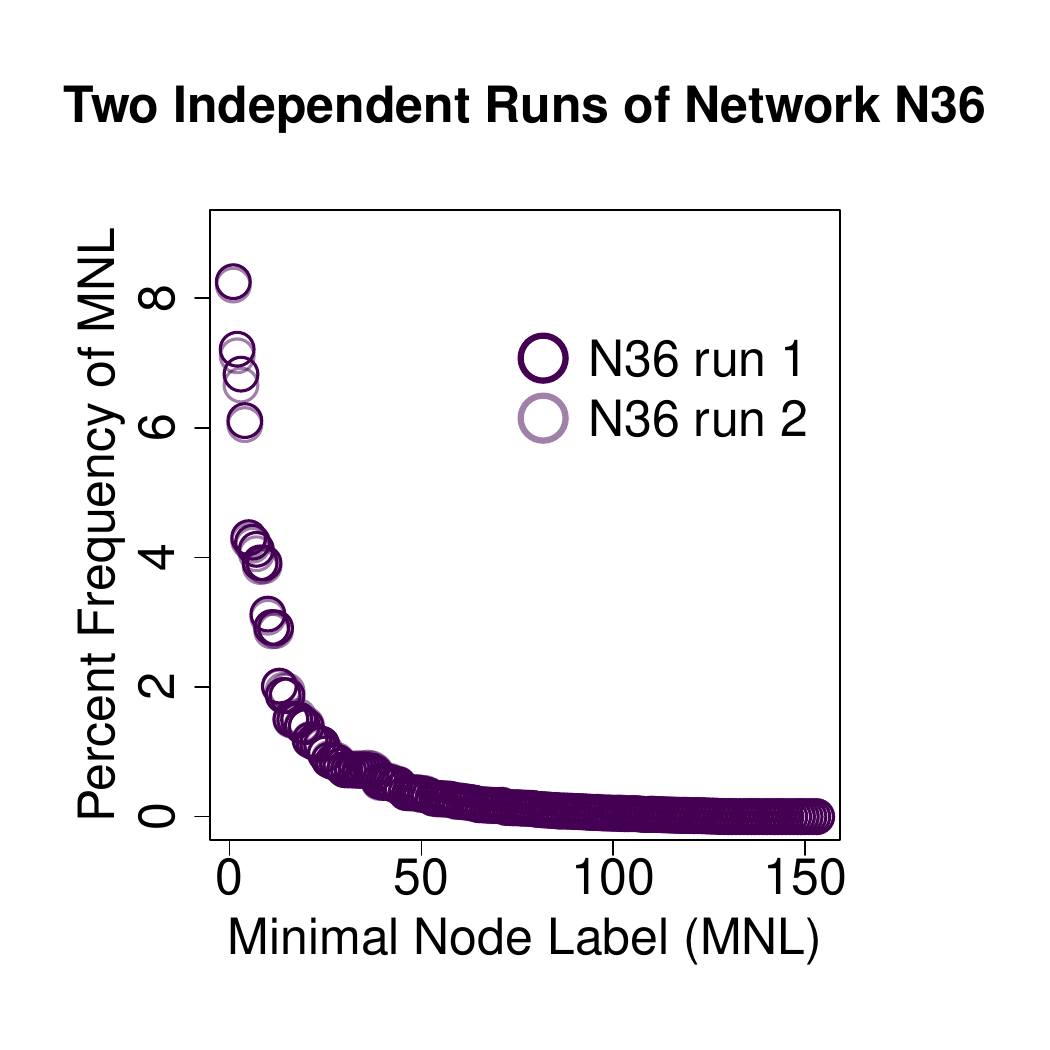}
}

\put(250,400){{\bf \large B}}
\put(250,260){ 
\includegraphics[scale=0.5]{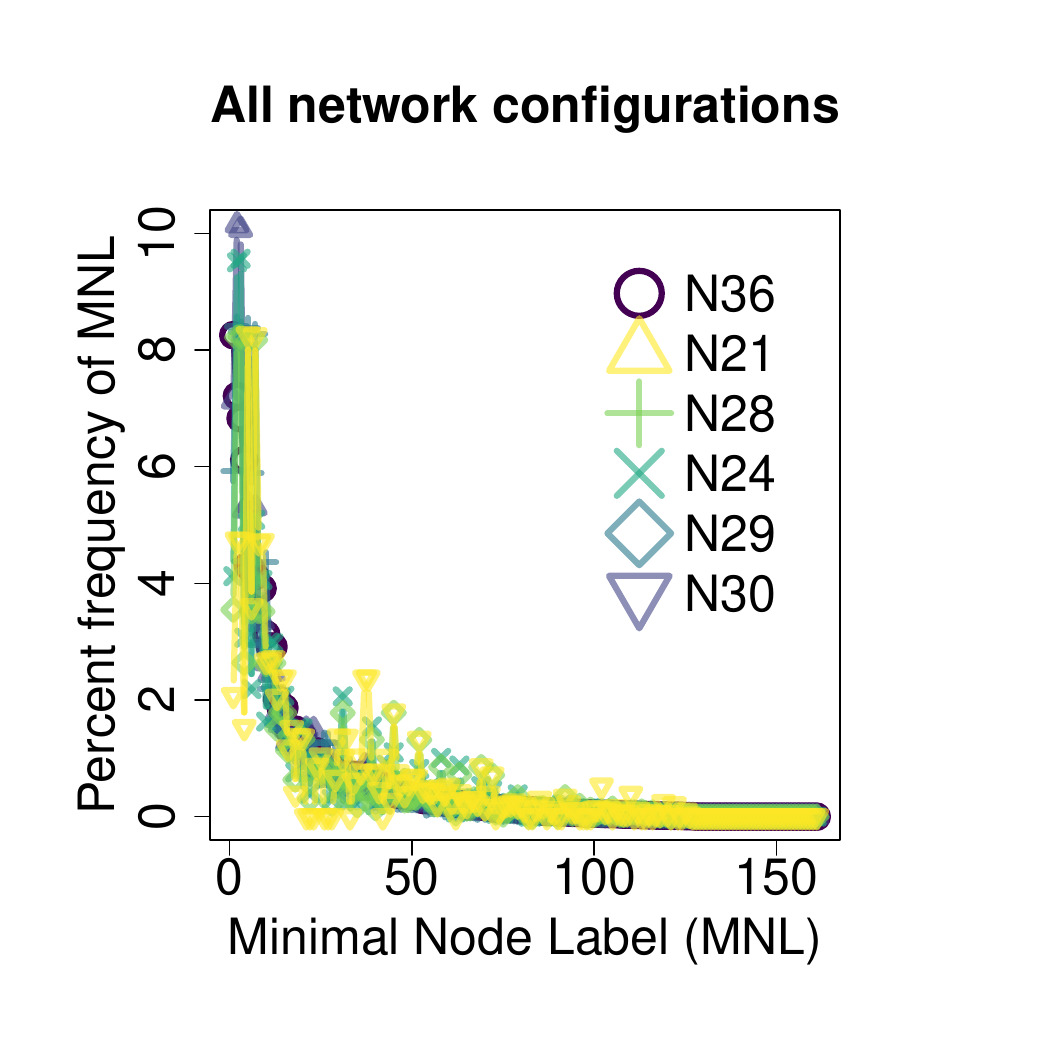}
}

\put(0,200){{\bf \large C}}
\put(10,10){ 
\includegraphics[scale=0.5]{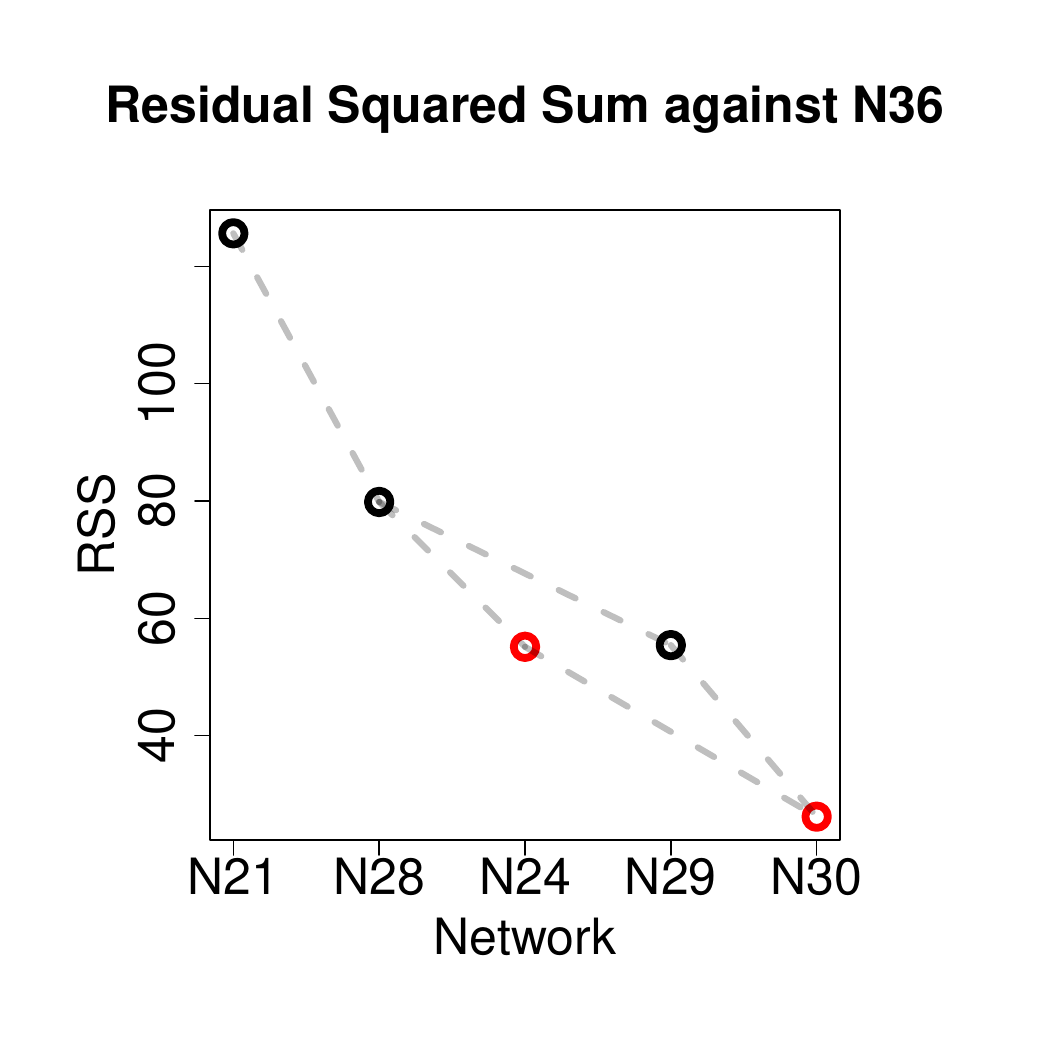}
}

\put(250,200){{\bf \large D}}
\put(250,10){ 
\includegraphics[scale=0.5]{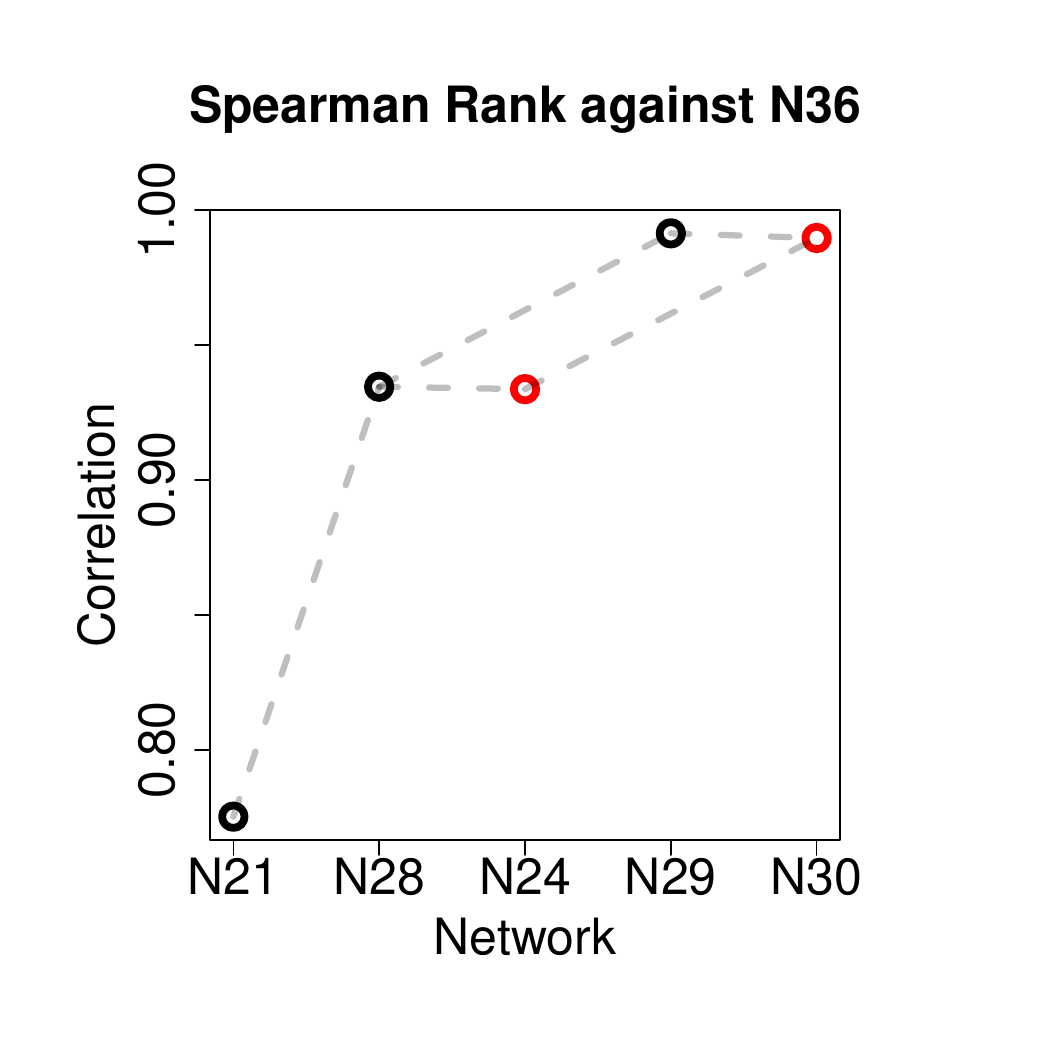}
}
\end{picture}
\caption{\footnotesize {\bf Panel A}: Multiple runs for network N36 using 30,000 randomly sampled parameter regions resulted in almost identical frequencies of minimal node labels (MNLs). {\bf Panel B}: The $161$ MNLs (based on 30,000 randomly sampled parameter regions) for networks N36, N21, N28, N24, N29, and N30.  {\bf Panel C}: The sum of the differences squared or residual squared sum (RSS) of the frequency of minimal nodes for each of the networks against the full network N36. {\bf Panel D}: The Spearman rank correlations for each of the networks against the full network N36. The dashed lines in panels C and D indicate the subnetwork structure: a dashed line from a left point to a right point indicates that the network corresponding to the left point is a subnetwork of the network corresponding to the right point. Black points indicate that the corresponding network has no interspecific competition between anemonefish species; red points indicate at least one anemonefish species competes against another.}
\label{fig:spearman_rss}
\end{figure}

Following the steps explained above, for ${\bf N36}$, the parameter space is $\R^{88}_+$ and is decomposed into more than $10^{32}$ distinct regions. Although the computation cost of determining an MG for a single region is just a few ($50$) seconds, it is clearly impossible to analyze the MG for all parameter regions. We therefore turn to random sampling. Surprisingly  (though consistent with what was observed in the predator-prey example), extremely sparse sampling seems to suffice. In particular, as is shown in Figure \ref{fig:spearman_rss}A for ${\bf N36}$ the percent frequency plots of the MNLs do not change when based on random samples of $30,000$ nodes in the parameter graph (see Figure \ref{fig:spearman_rss}A for ${\bf N36}$).

Not surprisingly, the percent frequency of the MNLs varied over each of the $5$ different networks (see Figure \ref{fig:spearman_rss}B).
However, the more similar to {\bf N36} the subnetwork is, the more similar the minimal node frequencies, as evidenced by the residual squared sums (RSS) and Spearman Rank correlations (see Figure \ref{fig:spearman_rss}C,D). 
Note that, roughly, more than $80\%$ of MNLs were rare (i.e., occurred with less than $1\%$ frequency; see Figure \ref{fig:spearman_rss}B). 

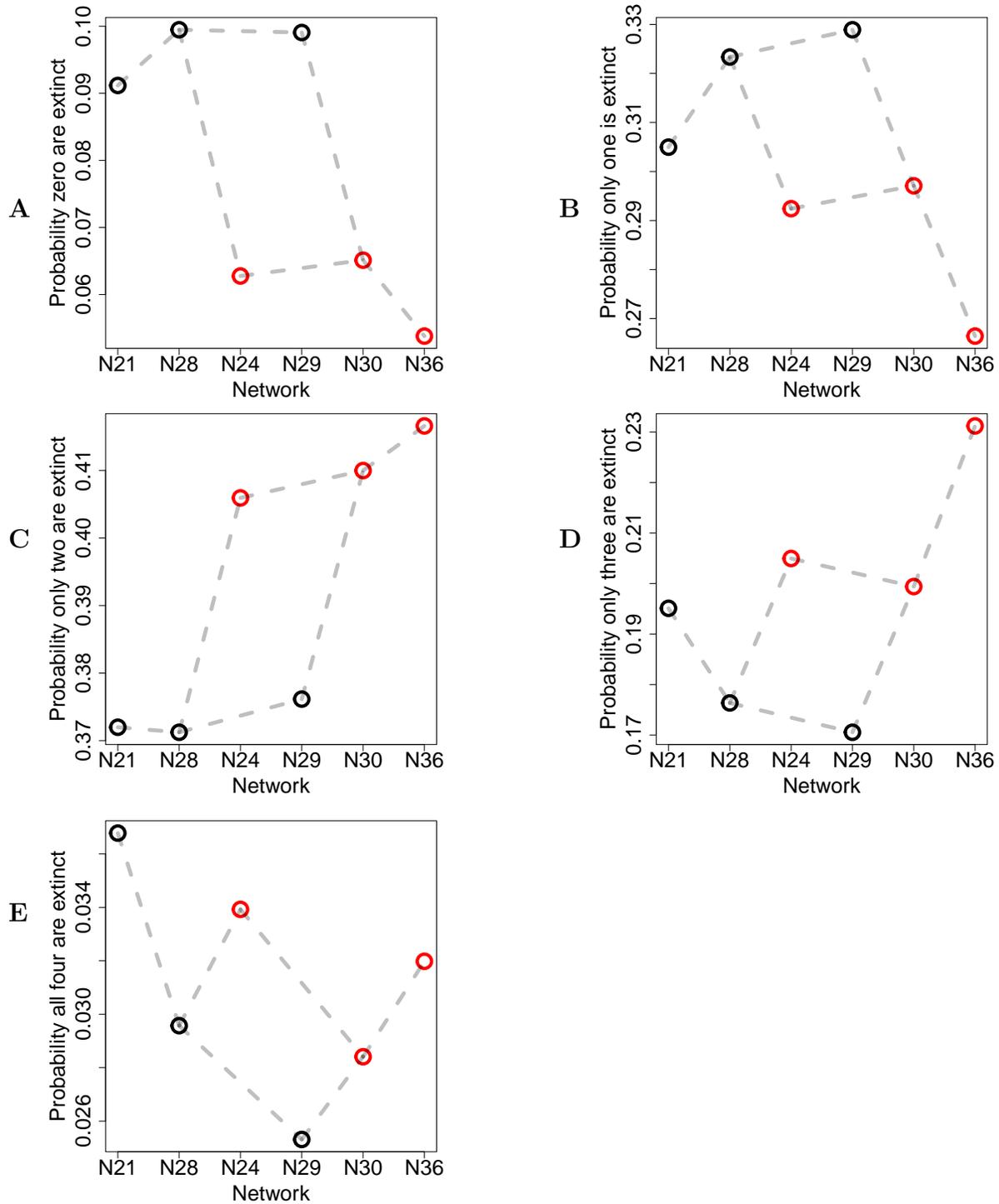
\begin{figure}[!htbp]
\begin{picture}(400,520)

\put(60,560){\large \bf Probability that only $0 \leq n \leq 4$ anemonefish species are extinct}

\put(0,450){{\large \bf A}}
\put(0,350){
\includegraphics[scale=0.4]{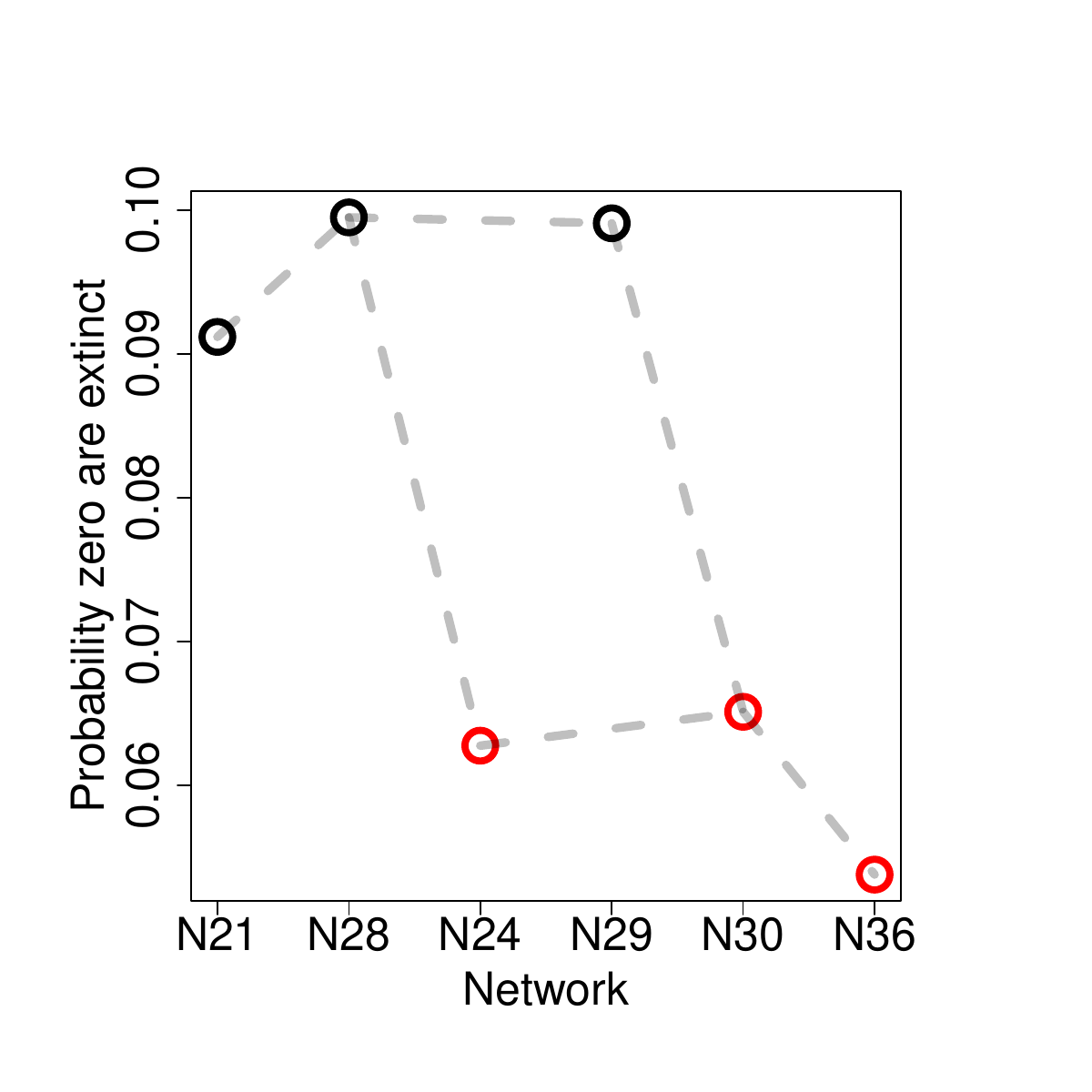}
}

\put(250,450){{\large \bf B}}
\put(250,350){
\includegraphics[scale=0.4]{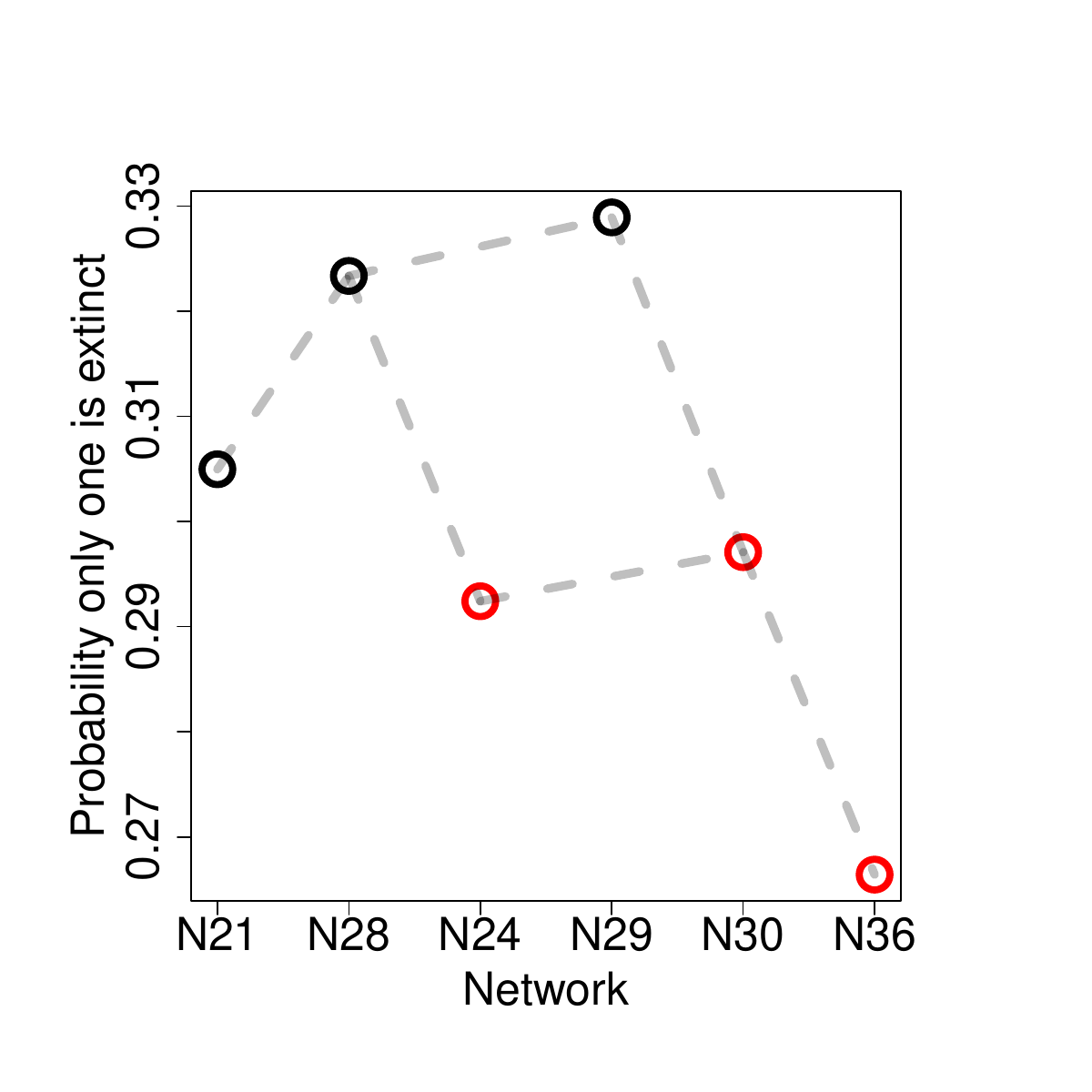}
}

\put(0,300){{\large \bf C}}
\put(0,170){
\includegraphics[scale=0.4]{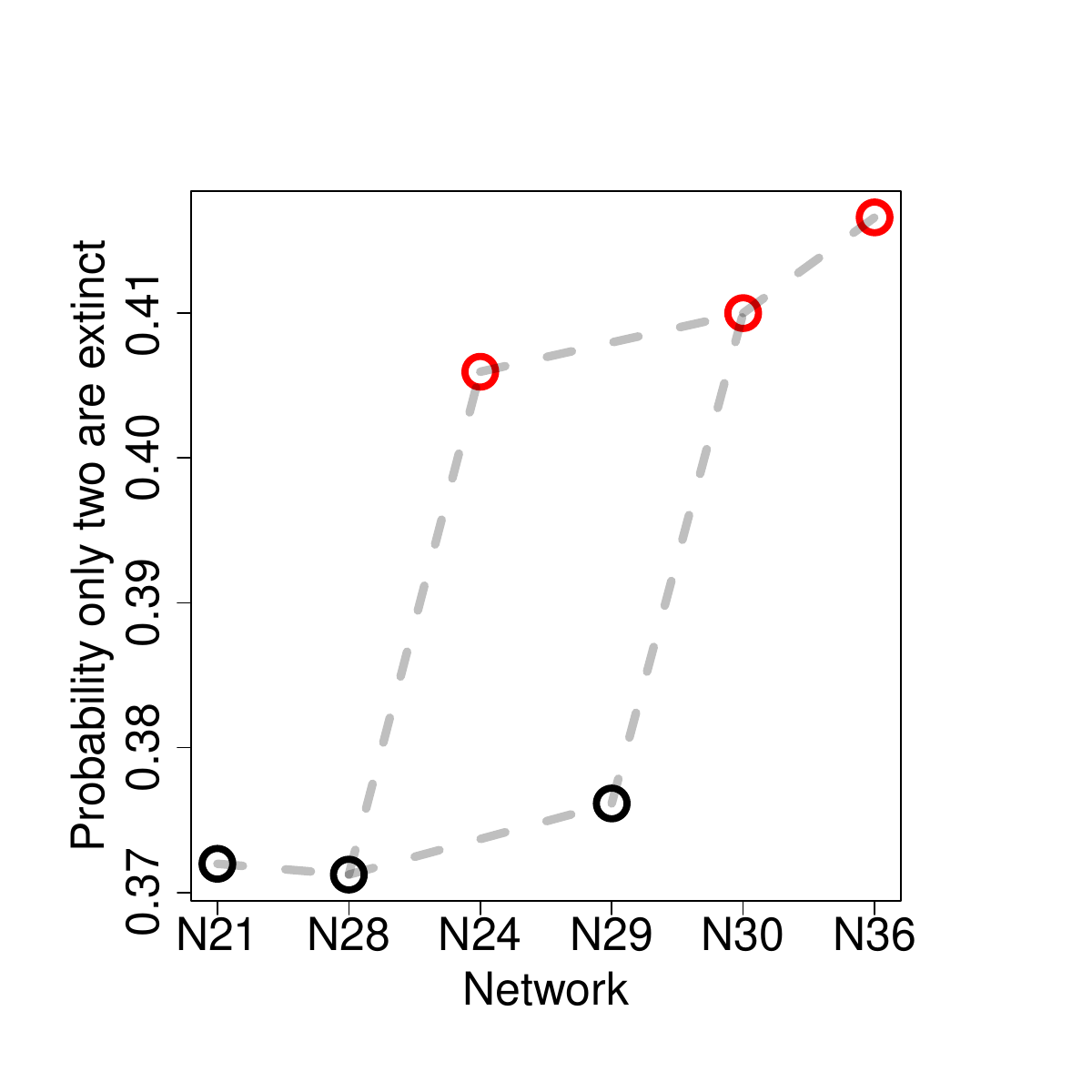}
}

\put(250,300){{\large \bf D}}
\put(250,170){
\includegraphics[scale=0.4]{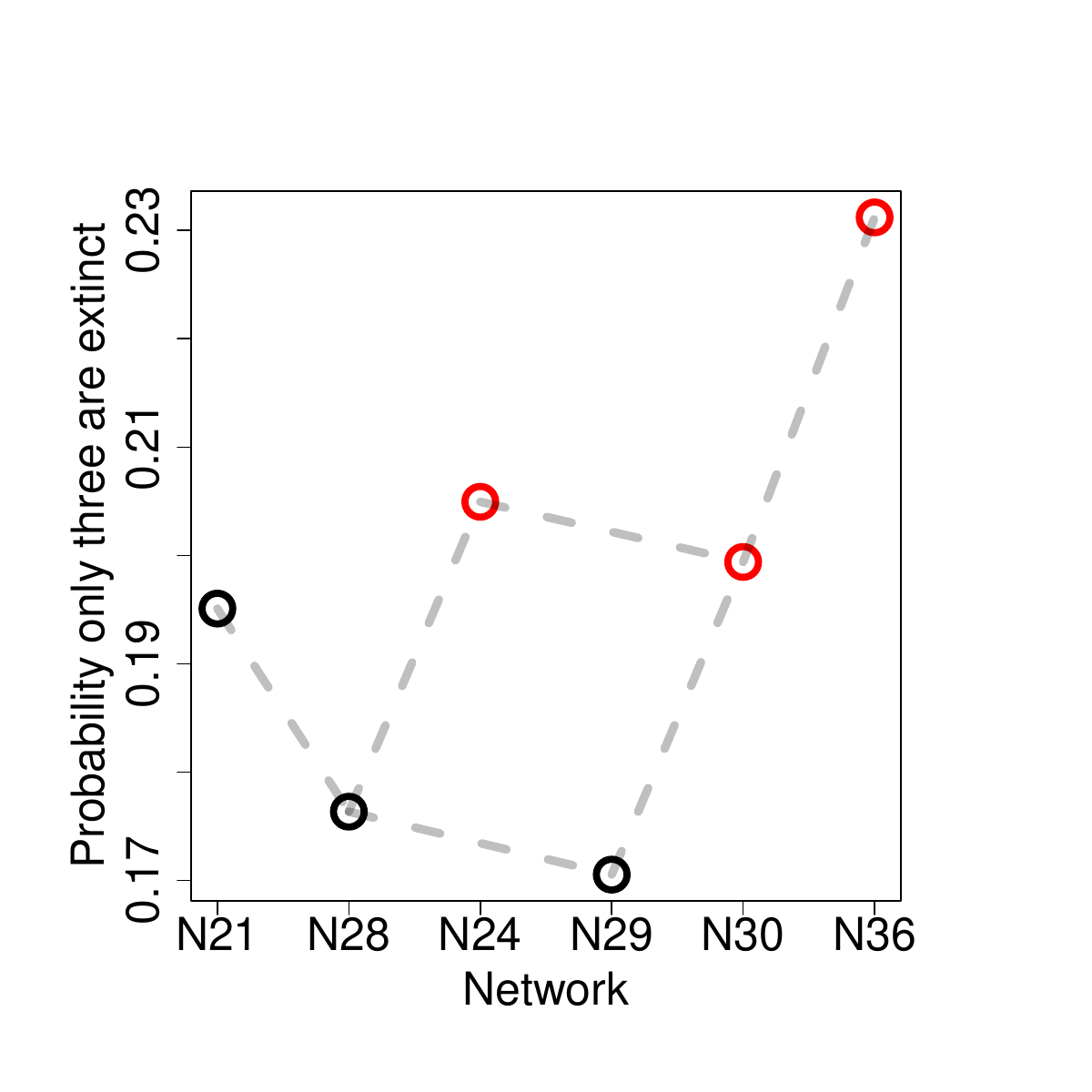}
}

\put(0,130){{\large \bf E}}
\put(0,-15){
\includegraphics[scale=0.4]{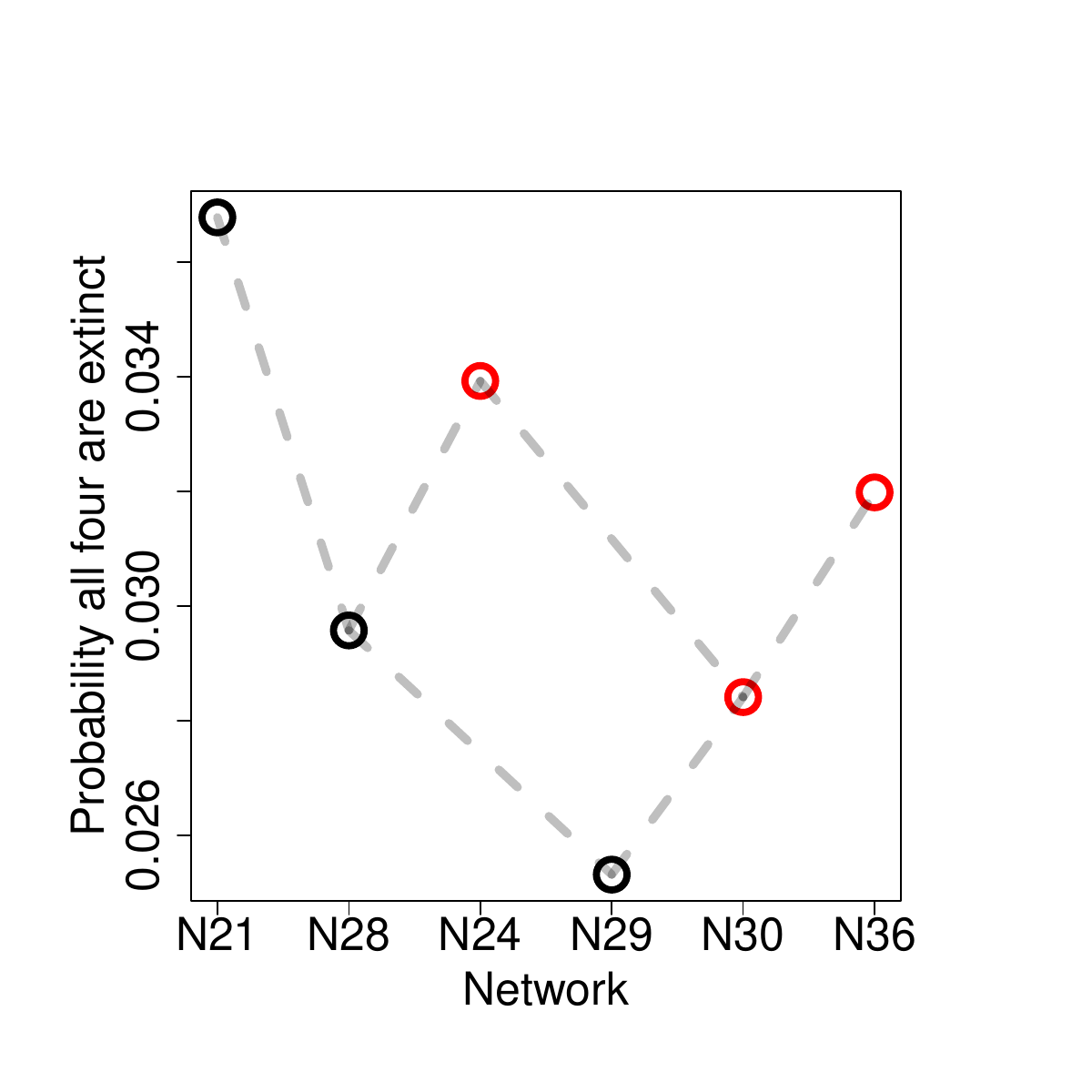}
}
\end{picture}
\caption{\footnotesize Percent frequency of  $\minnode$ labels corresponding to different long-term community dynamics. {\bf Panel A}: Percent frequency of locally attracting cells (community density regions) where all anemonefish species were present. {\bf Panel B}: Percent frequency of locally attracting community density regions in which one and only one anemonefish species went extinct. {\bf Panel C}: Percent frequency in which strictly two anemonefish species went extinct. {\bf Panel D}: Percent frequency in which strictly three anemonefish species went extinct. {\bf Panel E}: Percent frequency in which all four anemonefish species were extinct. A dashed line between two points indicates that the network corresponding to the left point is a subnetwork of the network corresponding to the right point. A point is red if at least one anemonefish species was in competition with another in that network configuration and is black otherwise.}
\label{fig:functional_species_probabilities}
\end{figure}

In locally attracting density regions (cells) it is rare for all anemonefish species to be extinct ($\approx 3.2\%$; see Figure \ref{fig:functional_species_probabilities}E), and it is more probable that one species ($\approx 23\%$; see Figure \ref{fig:functional_species_probabilities}D), two species ($\approx 42\%$; see Figure \ref{fig:functional_species_probabilities}C), or three species ($\approx 27\%$; see Figure \ref{fig:functional_species_probabilities}B) persist; an outcome in which all species are present is quite unlikely ($\approx 5\%$; see Figure \ref{fig:functional_species_probabilities}A). As a reminder, note that the final state for the community may depend on the starting density region, as MGs for our interaction networks often exhibit multistability, i.e., more than one minimal node ($\approx 96\%$ in {\bf N36}; see Figure \ref{fig:functional_species_probabilities}F). 
If one were to choose a  locally attracting density region at random, then the result will most likely be one in which two anemonefish species are present, although initializing the community sufficiently far away from this attractor may lead to another long-term outcome. 

Finally, we found that subnetworks that had competition between at least two of the anemonefish species ({\bf N30} and {\bf N24}) to be the closest to network configuration {\bf N36} (see Figure \ref{fig:functional_species_probabilities}A-D, red points and lines), thus highlighting the importance of competition for the community. The one exception was that configuration {\bf N28} (no competition between any anemonefish species) was marginally closer to configuration {\bf N36} than configuration {\bf N30} in the percentage of minimal nodes with full anemonefish extinction (see Figure \ref{fig:functional_species_probabilities}E).

\section{Discussion} 

$\wendy$ is an ecologically focused wrapper for an efficient computational platform, DSGRN, that takes as input an interaction network, constructs a model (and its phase space and parameter space), and outputs an explicit finite decomposition of the parameter space together with a combinatorial characterization of the dynamics in the form of a MG associated with each region of parameter space. This implies that, in principle, it is possible to compute \emph{all} possible robust dynamical structures.
$\wendy$, specifically, ensures that the internal model captures classical intra- and inter-species interactions (mutualism, competition, and predation) over an unbounded region of parameters, identifying fundamental biological features such as extinction. 
Minimal nodes in the MGs identify stable long-term community dynamics.
$\wendy$ extracts this information via MNLs that can be used for statistical analysis of expected community behavior over regions of this parameter space. In contrast, more traditional modeling approaches (e.g., ODEs, where solutions correspond to individual trajectories), provide much finer information but at a prohibitive computational cost: for example, creating a fine sampling of both parameter space and density space, computing trajectories for each such sample, and then identifying and organizing the simulated dynamics, but without guarantees of identification of all relevant phenomena.

Furthermore, as shown with the case study of the anemones and anemonefish, $\wendy$ allows for transparent incorporation of modeling assumptions and experimental observations to compute potential dynamics. Once the interaction network is established, computationally tractable outputs from $\wendy$ can be used to identify biological relevant information. For our case study, our framework predicts that significant perturbations to the anemones and anemonefish communities will lead to the loss of one or more species, but rarely lead to the loss of all species of a functional group, e.g, all anemonefish. 
This result thus provides information about the community. Nevertheless, the loss of one species could greatly alter community dynamics: for example, the extinction of a generalist anemonefish species (e.g., \emph{Amphiprion clarkii}) would necessarily remove any mutualistic benefit (e.g., protection from specialized fish predators) that the host anemone species receive from it \citep{schmitt2003mutualism}. $\wendy$ can also help assess the consequences of other factors that may be absent in the empirical data. For example, these anemone populations could drastically decline due to predation. This would directly decrease the number of available safe habitats for anemonefish species, and diminish their chances of survival (anemonefish are highly dependent on anemones protecting them and their offspring). This could increase intraspecific and interspecific competition between the remaining anemonefish species over the declining common resource (habitat space) and lead to the competitive exclusion of another anemonefish species \citep{hardin1960competitive}. 
Indeed, as evidenced by our results, the existence of competition alone can increase the number of density regions in which exactly two or exactly three anemonefish species are excluded. 

To put these findings into perspective, it should be noted that the field experimentalists observed a system in which all species were present (and would, otherwise, not be present in the initial network configuration). With respect to $\wendy$'s outputs, the most reasonable assumption is that the observed dynamics are associated with a minimal Morse node that supports all species, and thus the existing ecosystem is rare. In particular, a large perturbation to the system and the community's population densities could easily lead to an alternative stable state for which it is expected that fewer species will be supported. Alternatively, changes in parameters (e.g., net growth rates or growth rates associated with the interactions of species) will also likely lead to fewer species being supported.  In particular, if a strong enough disturbance occurs in the ecosystem and alters community population densities, there is a higher potential that the community will not recover altogether; instead, this perturbation could put some species on the track to extinction. Thus, our framework can provide the necessary information for decision-makers to assess the direct and indirect consequences of environmental or ecosystem changes and management strategies.

We conclude with a few comments indicating how applications of $\wendy$ can be strengthed. The most obvious bottleneck is the number of elements in the decomposition of parameter space; the number of parameter regions grows extremely rapidly as a function of the complexity of the trophic network \citep{kepley2021computing}. However, note that the computations performed in this paper were done without any assumptions on parameter values. Constraining the parameter space with even just relative information (e.g., is the impact of intraspecific competition greater than or less than the impact of interspecific competition or mutualism?) would dramatically reduce the number of biologically relevant nodes in the parameter graph. This bottleneck thus helps identify gaps in the information available for a focal system, which could be bridged with multidisciplinary efforts. As noted above, a reasonable assumption is that within the $\wendy$ model the current ecosystem is operating within a parameter region (a node in the parameter graph) that supports co-existence of all species of anemones and anemonefishes. Ideally, we would like to be able to identify all nodes with this property and understand how these nodes are embedded in the full parameter graph. A single large cluster would suggest that the ecosystem is robust with respect to perturbations, whereas a slender fragmented embedding suggests that a wide variety of perturbations could lead to the elimination of some of the species. For small networks, explorations of this type are possible. However, given the size of the parameter graph for interaction networks of ecological relevance, the capability to perform such explorations requires new (both theoretical and computational) mathematical developments.  

Finally,  $\wendy$ does not currently provide a way to incorporate spatial dynamics, e.g., migration.
In principle, one can re-interpret species and the strength of interactions between species as patches within a matrix of landscapes and include terms that indicate strength of migration from one patch to another. However, to provide the user with tractable tools to do this in a computationally efficient way requires nontrivial modifications to the DSGRN code base and $\wendy$ wrapper.  

\section{Methods}

We extended and modified Dynamic Signatures Generated by Regulatory Networks (DSGRN) to create $\wendy$ (see the Appendix). In summary, we first removed the restriction that DSGRN does not allow multiple terms involving one species (see Eq.(4) of Section \ref{sec:2species}). Second, we changed the prescription of the growth-rate function (in DSGRN terms, the interaction function) to be a summation of the interactions between a species and another. Finally, we introduced a small modification to the state transition graph to allow for the detection of unstable equilibria representing the extinction of one or more species.  

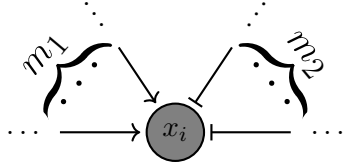
\begin{figure}[!htbp]
\begin{picture}(400,50)
\put(150,0){

\begin{tikzpicture}[thick, scale=0.9]

\node[circle, fill=gray, draw] (1) at (0,0) {$x_i$};
\node[circle, fill=white, draw=none] (2) at (-2,0) {};
\node[circle, fill=white, draw = none] (3) at (-1,1.5) {};
\node[circle, fill=white, rotate = 50, draw = none] (4) at (-1.75,1) {\huge $\overbrace{\ldots}^{m_1}$ };

\node[circle, fill=white, draw = none] (5) at (2,0) {};
\node[circle, fill=white, draw = none] (6) at (1,1.5) {};
\node[circle, fill=white, rotate = -50, draw = none] (7) at (1.75,1) {\huge $\overbrace{\ldots}^{m_2}$ };

\draw[->, shorten <= 2pt, shorten >= 2pt] (2) to (1);
\draw[->, shorten <= 2pt, shorten >= 2pt] (3) to (1);

\draw[-|, shorten <= 2pt, shorten >= 2pt] (5) to (1);
\draw[-|, shorten <= 2pt, shorten >= 2pt] (6) to (1);

\node[circle, fill=white, draw=none] (8) at (-2.25,0) {$\ldots$};
\node[circle, fill=white, draw=none] (9) at (2.25,0) {$\ldots$};
\node[circle, fill=white, rotate=55, draw=none] (10) at (1.2,1.75) {$\ldots$};
\node[circle, fill=white, rotate=125, draw=none] (11) at (-1.2,1.75) {$\ldots$};
\end{tikzpicture}
}
\end{picture}
\caption{\footnotesize A representation of part of a network configuration in which species $i$ has $m_1$ beneficial interactions with $m_1$ other species and $m_2$ detrimental interactions with $m_2$ other species. Beneficial and detrimental interactions are denoted with a pointed arrow $\rightarrow$ and blunted arrow $\dashv$, respectively. Note that the detrimental interaction can come from itself (as with intraspecific competition).}
\label{fig:arbitrary_numbered_incoming_edges_to_species_i}
\end{figure}

In general, we defined species $i$ growth rate function as follows. Respectively, a pointed or blunted edge from species $j$ node to species $i$ node introduces an interaction term $\sigma_{i,i}^{(k)}\sigma_{i,j}^{(0)}$ that promotes or inhibits the growth rate $r_i$ of species $i$ if species $j$ and species $i$ densities have surpassed arbitrary density thresholds $\theta_{i,j}^{(0)}$ and $\theta_{i,i}^{(k)}$ (otherwise, the impact is zero); here,

\begin{align}
\sigma_{i,j}^{(l)}(x_j) =
\begin{cases}
0, & \textrm{if } x_{j} < \theta_{i,j}^{(l)}\\
\delta_{i,j}^{(l)}, & \textrm{if } x_{j} > \theta_{i,j}^{(l)}.
\end{cases}
\end{align}
In particular, assume species $i$ node is grey-filled, i.e., species $i$ has a positive background growth rate $\gamma_i$, it has $m_1$ incoming pointed edges from $m_1$ species different from species $i$; and it has $m_2$ incoming blunted edges from $m_2$ species that are not species $i$ (see Figure \ref{fig:arbitrary_numbered_incoming_edges_to_species_i}). Then

\begin{align}
\label{eq:growth_rate}
r_i(x)  = \gamma_i x_i + \sum_{p=1}^{m_1} \sigma_{i,i}^{(k_p)} \sigma_{i,i_p}^{(0)} - \sum_{n = 1}^{m_2} \sigma_{i,i}^{(k_{n} )} \sigma_{i,i_{n}}^{(0)}.
\end{align}
If, on the other hand, species $i$ has a negative background growth rate (i.e., its node is white-filled), then $r_i$ has a $-\gamma_i$ instead in the above equation. We note that if the target and source node of a directed arrow are the same, i.e., $i = j$, then the term added or subtracted (depending on whether the directed arrow in pointed or blunt, respectively) in Eq.\eqref{eq:growth_rate} is just
\begin{align}
\sigma_{i,i}^{(0)}(x_i) =
\begin{cases}
0, & \textrm{if } x_{i} < \theta_{i,i}^{(0)}\\
\delta_{i,i}^{(0)}, & \textrm{if } x_{i} > \theta_{i,i}^{(0)}.
\end{cases}
\end{align}
In particular, we attain the first subtracted term that arises for the prey species' growth rate equations in equations Eq.\eqref{eq:rate_change} and Eq.\eqref{eqtn:preygrowthrate_in_predprey}.

\section{Appendix: DSGRN Extension to Ecological Networks}
\label{sec:DSGRN_extension}

The Dynamic Signatures Generated by Regulatory Networks (DSGRN) software \cite{cummins2016combinatorial, cummins:gameiro:gedeon:kepley:mischaikow:zhang} takes a regulatory network and an interaction function (a function describing how the inputs to a node interact \cite{kepley2021computing, cummins:gameiro:gedeon:kepley:mischaikow:zhang}) as input, creates an explicit finite decomposition of the parameter space, and for each region of the parameter space computes a combinatorial description of the global dynamics. The explicit decomposition of parameter space is given by a finite collection of semi-algebraic sets (represented by explicit inequalities involving the parameters), with the property that the combinatorial dynamics is constant for all parameter values within each of these semi-algebraic sets. The decomposition of parameter space is organized by the \emph{parameter graph}, where the vertices represent the semi-algebraic sets defining the regions in parameter space and an edge between two vertices indicates that the corresponding regions are co-dimension $1$ neighbors. The combinatorial description of the global dynamics is obtained via a cell complex $\cX$ decomposition of the phase space and the construction of a multi-valued map $\cF$ mapping cells of $\cX$ into a collection of cells of $\cX$. The multi-valued map $\cF$ is represented by a directed graph (digraph) called the \emph{state transition graph} (STD). The state transition graph represents a combinatorial discretization of the system and is used to extract a combinatorial description of global dynamics via graph algorithms. More precisely, regions exhibiting recurrent dynamics are represented by the nontrivial strongly connected components of $\cF$, which can be identified in linear time \cite{cormen:leiserson:rivest:stein}.

In this section we present an extension and modification to DSGRN to compute the dynamics of ecological interaction networks. We only present the details of the proposed modifications to DSGRN needed for this extension and refer the reader to \cite{cummins2016combinatorial, cummins:gameiro:gedeon:kepley:mischaikow:zhang} for an in-depth description of the original DSGRN and its most recent version. The extensions to DSGRN to ecological networks presented in this paper are: (1) The DSGRN software does not allow for multiple directed edges between two vertices of the network, that is, given two vertices $v_1$ and $v_2$ at most one edge from $v_1$ to $v_2$ is allowed. Here we remove this restriction and allow for multiple edges from $v_1$ to $v_2$; (2) We introduce a new type of \emph{activating function} modeling how one node of the network affects its target node; (3) We introduce a new class of \emph{interaction functions} to model ecological networks; (4) Finally, we introduce a small modification to the \emph{state transition graph} (STG) to allow for the detection of unstable equilibria representing the extinction of one or more species.

\subsection{Ecological Networks}

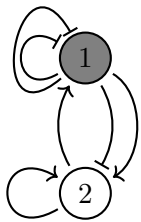
\begin{figure}[!htbp]
\centering
\begin{picture}(50,80)
\put(-10,-10){
\begin{tikzpicture}[thick, scale=0.9]
\node[circle, fill=gray, draw] (1) at (0,2) {$1$};
\node[circle, fill=white, draw] (2) at (0,0) {$2$};
\draw[-|, shorten <= 2pt, shorten >= 2pt] (1) to [bend left] (2);
\draw[->, shorten <= 2pt, shorten >= 2pt] (2) to [bend left] (1);
\draw[->, shorten <= 2pt, shorten >= 2pt] (1) to [out=-30, in=30] (2);
\draw[-|, shorten <= 2pt, shorten >= 2pt] (1) to [out=210, in=150, looseness=6] (1);
\draw[-|, shorten <= 2pt, shorten >= 1pt] (1) to [out=220, in=120, looseness=7] (1);
\draw[->, shorten <= 2pt, shorten >= 2pt] (2) to [out=210, in=150, looseness=8] (2);
\end{tikzpicture}
}
\end{picture}
\caption{Network with two nodes, where node $1$ is positive and node $2$ is negative. Positive edges are indicated with a pointy arrow ($\to$) and negative edges with a blunt arrow ($\dashv$).}
\label{fig:network_example}
\end{figure}

One of the inputs to the DSGRN extension proposed here is a network, represented as directed graph with annotated vertices and edges. 
The nodes are indexed $1$ to $N$. 
A node $n$ is annotated with  $+$ (or  $-$) to indicate that the quantity $x_n$ associated with it has an intrinsic growth (or decay) rate. 
We denote a node by $(n, +)$ or $(n, -)$ when we want to explicitly indicate the sign of the node. 
Otherwise, we simply denote it by $n$. 
An edge from node $i$ to node $j$ is annotated with  $+$ (or  $-$) to indicate that the quantity $x_i$ increases (or decreases) the rate of production of $x_j$. 
Edges from $i$ to $j$ need not be unique, hence an edge is further annotated with an index $k \in \{ 0, 1, 2, \ldots \}$, called the \emph{instance} of the edge. 
An edge from $i$ to $j$ is denoted by $(i, j, k, +)$ or $(i, j, k, -)$ when we want to explicitly indicate the sign and the instance of the edge. Otherwise, we denote a positive edge by $i \to j$ and a negative edge by $i \dashv j$ when we want to indicate just the sign, or simply by $(i, j)$ if we do not wish to explicitly indicate the sign and instance. Pictorially, we indicate positive vertices by shaded nodes, negative nodes by non-shaded nodes, positive edges by $\to$, and negative edges by $\dashv$ (see Figure~\ref{fig:network_example}).

\subsection{Activating function}

To each edge $(m, n, k, \pm)$ we associate an \emph{activating function}
\[
\sigma^{(k)}_{n,m}(x_m) :=
\begin{cases}
0,                  & \text{if } x_m < \theta^{(k)}_{n,m} \\
\delta^{(k)}_{n,m}, & \text{if } x_m > \theta^{(k)}_{n,m}
\end{cases}
\]
which is used to represent how $x_m$ affects $x_n$ via the edge $(m, n, k, \pm)$. Notice that the sign of the edge does not play a role in the definition of $\sigma^{(k)}_{n,m}$.

\subsection{Interaction functions and the rate of change expression}

To each node $n$ in the network we associate a positive parameter $\gamma_n$, which represents the intrinsic decay or growth rate of $x_n$. The \emph{rate of change} of $x_n$ is given by a function of the form
\begin{equation}
\label{eq:rate_of_change}
\pm \gamma_n x_n + \Lambda_n(x)
\end{equation}
where the sign of the $\gamma_n$ term is the same as the sign of the node, that is, the term is $-\gamma_n x_n$ if the node $n$ is a negative node $(n, -)$ and it is $+\gamma_n x_n$ if the node $n$ is a positive node $(n, +)$. To describe the allowable forms of $\Lambda_n$, we need the following definition.

\begin{definition}
\label{def:interaction_function}
An \emph{interaction function} of order $J$ is a polynomial in $J$ variables $z = (z_1, \ldots, z_J)$ of the form 
\[
f(z) := \sum_{j=1}^q f_j(z)
\]
where each term has the form 
\[
f_j(z) = \pm \prod_{i\in I_j}z_i
\]
and the indexing sets $\setof{I_j\mid 1\leq j\leq q}$ form a partition of $\setof{1,\ldots, J}$.
\end{definition}

Consider a node $n$ with $J$ in-edges $(m_1, n, k_1, \pm), (m_2, n, k_2, \pm), \ldots, (m_J, n, k_J, \pm)$. We assume that the function $\Lambda_n$ has the form
\[
\Lambda_n(x_{m_1}, x_{m_2}, \ldots, x_{m_J}) = f(\sigma^{(k_1)}_{n, m_1}(x_{m_1}), \sigma^{(k_2)}_{n, m_2}(x_{m_2}), \ldots, \sigma^{(k_J)}_{n, m_J}(x_{m_J})),
\]
where $f$ is an interaction function of order $J$ with indexing sets satisfying the condition that if two indices $i_1$ and $i_2$ are in the same indexing set $I_j$, then the corresponding edges $(m_{i_1}, n, k_{i_1}, \pm)$ and $(m_{i_2}, n, k_{i_2}, \pm)$ must have the same sign and this is the sign of the term $f_j$ in the interaction function. The final condition in the interaction function $f$ defining $\Lambda_n$ is that at least one of the terms $f_j$ must have the opposite sign of the sign of the node $n$.

We sometimes make a slight abuse of notation and refer to the polynomial expression of the interaction function $f$ as its \emph{interaction type}. As an example, we may say that the interaction function $f(x) = x_1 x_2 - x_3$ is an interaction of type $x_1 x_2 - x_3$.

\begin{example}
\label{ex:rate_change_example}
Consider the network in Figure~\ref{fig:network_example}. The allowed interaction types for node $1$ are $-x_1 x_1 + x_2$ or $-x_1 -x_1 + x_2$ and for node $2$ are $-x_1 + x_1 x_2$ or $-x_1 + x_1 + x_2$. Notice that these are just algebraic expressions indicating the interaction types, and hence $-x_1 + x_1$ does not cancel out in the last expression. If we choose the interaction types $-x_1 x_1 + x_2$ and $-x_1 + x_1 + x_2$ for nodes $1$ and $2$, respectively, then the rate of change expressions for $x_1$ and $x_2$ are given by
\[
\begin{cases}
~~~\gamma_1 x_1 - \sigma^{(0)}_{1,1}(x_1) \sigma^{(1)}_{1,1}(x_1) + \sigma^{(0)}_{1,2}(x_2) & \\
-\gamma_2 x_2 - \sigma^{(0)}_{2,1}(x_1) + \sigma^{(1)}_{2,1}(x_1) + \sigma^{(0)}_{2,2}(x_2) &
\end{cases}
\]
\end{example}

\subsection{Parameter space decomposition}

To determine the state transition graph it is necessary to determine the signs of \eqref{eq:rate_of_change} when evaluated at the thresholds $\theta^{(k)}_{*, n}$. The signs of \eqref{eq:rate_of_change} can be determined in terms of the parameters and form a decomposition of the parameter space into semi-algebraic sets (defined in terms of inequalities involving the parameters). This decomposition of the parameter space is represented by the parameter graph (see \cite{cummins2016combinatorial} for details).

Note that it is sufficient to consider the case $-\gamma_n x_n + \Lambda_n(x)$ since the other case can be reduced to this by flipping the sign. The signs of $-\gamma_n x_n + \Lambda_n(x)$ can be determined using the same methods used to compute the parameter decomposition in DSGRN \cite{cummins2016combinatorial, cummins:gameiro:gedeon:kepley:mischaikow:zhang, kepley2021computing}. One can also determine the parameter space decomposition from a corresponding DSGRN parameter decomposition as described in the following.

Let $\Lambda_n$ be given by an interaction function $f$ with $q$ terms determined by the indexing sets $I_j$ for $1 \leq j \leq q$ and let $\prod_{i \in I_j} \sigma^{(k_i)}_{n, m_i}(x_{m_i})$ denote that the term of $\Lambda_n$ corresponding to $I_j$. Denoting
\[
\tilde{\delta}_{n,j} := \prod_{i \in I_j} \delta^{(k_i)}_{n, m_i}
\]
we note that this term can take only the values $0$ or $\tilde{\delta}_{n,j}$. It follows that $\Lambda_n(x)$ can take only the values
\begin{equation}
\label{eq:rate_change_values}
\left\{ \alpha_1 \tilde{\delta}_{n,1} + \cdots + \alpha_q \tilde{\delta}_{n,q} \mid \alpha_j \in \{ 0, \pm 1 \} \right\}
\end{equation}
where the sign of $\alpha_j$ is the sign of the term $f_j$ in the interaction function. The parameter space decomposition can be determined from the set of all total orders of the values of $\Lambda_n(x)$ \cite{kepley2021computing}, which in turn can be determined from the set of all orderings of \eqref{eq:rate_change_values}. We refer to the values of $\Lambda_n(x)$ as the \emph{input polynomials} to node $n$ and the values in \eqref{eq:rate_change_values} as the \emph{reduced input polynomials} to node $n$.

\begin{example}
\label{ex:Lambda_values}
The function $\Lambda_1 (x) = - \sigma^{(0)}_{1,1}(x_1) \sigma^{(1)}_{1,1}(x_1) + \sigma^{(0)}_{1,2}(x_2)$ from Example~\ref{ex:rate_change_example} can take the values
\begin{align*}
p_0 & := -0 \cdot 0 + 0 = 0 & p_4 & := -0 \cdot 0 + \delta^{(0)}_{1,2} = \delta^{(0)}_{1,2} \\
p_1 & := -\delta^{(0)}_{1,1} \cdot 0 + 0 = 0 & p_5 & := -\delta^{(0)}_{1,1} \cdot 0 + \delta^{(0)}_{1,2} = \delta^{(0)}_{1,2} \\
p_2 & := -0 \cdot \delta^{(1)}_{1,1} + 0 = 0 & p_6 & := -0 \cdot \delta^{(1)}_{1,1} + \delta^{(0)}_{1,2} = \delta^{(0)}_{1,2} \\
p_3 & := -\delta^{(0)}_{1,1} \delta^{(1)}_{1,1} + 0 = -\delta^{(0)}_{1,1} \delta^{(1)}_{1,1} & p_7 & := -\delta^{(0)}_{1,1} \delta^{(1)}_{1,1} + \delta^{(0)}_{1,2}.
\end{align*}
These are the input polynomials to node $1$. The reduced input polynomials are
\[
\tilde{p}_0 := 0, \quad \tilde{p}_1 := -\tilde{\delta}_{1,1}, \quad \tilde{p}_2 := \tilde{\delta}_{1,2}, \quad \tilde{p}_3 := -\tilde{\delta}_{1,1} + \tilde{\delta}_{1,2},
\]
where $\tilde{\delta}_{1,1} := \delta^{(0)}_{1,1} \delta^{(1)}_{1,1}$ and $\tilde{\delta}_{1,2} := \delta^{(0)}_{1,2}$. The possible orders for these polynomials are
\[
\tilde{p}_1 < \tilde{p}_0 < \tilde{p}_3 < \tilde{p}_2 \quad \text{and} \quad \tilde{p}_1 < \tilde{p}_3 < \tilde{p}_0 < \tilde{p}_2.
\]
\end{example}

\begin{remark}
The only place where we use the fact that the low value of the activating function $\sigma^{(k)}_{n,m}$ is zero, instead of a non-zero value $\ell^{(k)}_{n,m}$ as in the original DSGRN, is to obtain the a set of reduced input polynomials that is smaller than the set of input polynomials when we have terms of size more than one in the activating function. This in turn makes it possible use the DSGRN parameter decomposition to derive the parameter decomposition for ecological networks interaction functions with a larger number of variables. Everything else would work exactly the same if we were to make the activating function have a non-zero low value.
\end{remark}

Next, we need to determine the set of all possible ordering of the reduced input polynomials in \eqref{eq:rate_change_values}. First note that we only need to consider the case where all signs in \eqref{eq:rate_change_values} are positive, since the set of all possible orderings in the general case can be determined from the orders of the corresponding terms where all the signs are made positive, simply by subtracting the negative terms from each of the orders of the set of positive terms as in the example below.

\begin{example}
Consider the reduced polynomials
\[
\tilde{p}_0 := 0, \quad \tilde{p}_1 := -\tilde{\delta}_{1,1}, \quad \tilde{p}_2 := \tilde{\delta}_{1,2}, \quad \tilde{p}_3 := -\tilde{\delta}_{1,1} + \tilde{\delta}_{1,2},
\]
from the previous example. For each $\tilde{p}_i$ we define a corresponding $p_i$ obtained from $\tilde{p}_i$ by flipping all negative signs to positive, that is, we define
\[
p_0 := 0, \quad p_1 := \tilde{\delta}_{1,1}, \quad p_2 := \tilde{\delta}_{1,2}, \quad p_3 := \tilde{\delta}_{1,1} + \tilde{\delta}_{1,2}.
\]
Now given an ordering of $\{ p_0, p_1, p_2, p_3 \}$ we can determine the corresponding order of $\{ \tilde{p}_0, \tilde{p}_1, \tilde{p}_2, \tilde{p}_3 \}$ by constructing the bijection between these two sets of polynomials obtained by subtracting from the $p_i$ all the negative terms in the $\tilde{p}_i$. Consider the order
\[
p_0 < p_1 < p_2 < p_3.
\]
Subtracting $-\tilde{\delta}_{1,1}$, the only negative term in the $\tilde{p}_i$, from the terms in this ordering we get
\[
-\tilde{\delta}_{1,1} < \tilde{\delta}_{1,1} -\tilde{\delta}_{1,1} < \tilde{\delta}_{1,2} -\tilde{\delta}_{1,1} < \tilde{\delta}_{1,1} + \tilde{\delta}_{1,2} -\tilde{\delta}_{1,1},
\]
which gives the ordering
\[
\tilde{p}_1 < \tilde{p}_0 < \tilde{p}_3 < \tilde{p}_2.
\]
Analogously the ordering $p_0 < p_2 < p_1 < p_3$ gives the ordering $\tilde{p}_1 < \tilde{p}_3 < \tilde{p}_0 < \tilde{p}_2$.
\end{example}

Finally note that when the terms in \eqref{eq:rate_change_values} all have positive signs, then the set of all possible orderings of these polynomials is precisely the set of all ordering of the original DSGRN input polynomials with a linear interaction function of order $q$, which have been previously determined for $q = 1, 2, \ldots, 6$ \cite{kepley2021computing}. Therefore we can derive the total orderings, and hence the parameter decomposition, from the original DSGRN parameter decompositions for all ecological networks with interaction functions with $6$ or less terms.

\subsection{State Transition Graph}

The \emph{state transition graph} in DSGRN is a digraph where the vertices are the cubical cells in the cell complex $\cX$ and the edges indicate the direction of the flow in the cell complex and are determined via the signs of \eqref{eq:rate_of_change}, which are encoded in the parameter graph. We determine the state transition graph for a given parameter node as it is done in the current version of DSGRN \cite{cummins:gameiro:gedeon:kepley:mischaikow:zhang}, except for the modification on when to assign a self-edge described below. Recall that in the current version of DSGRN \cite{cummins:gameiro:gedeon:kepley:mischaikow:zhang} we do a ``blow-up'' of thresholds corresponding to (both positive and negative) self-edges in the network, that is, we replace such thresholds with two thresholds (see \cite{gameiro:gedeon:kepley:mischaikow:2021} for details). Hence here we also do a blow-up of thresholds corresponding to self-edges in the network.

\subsubsection{Self-Edges in the State Transition Graph}

In the current version of DSGRN \cite{cummins:gameiro:gedeon:kepley:mischaikow:zhang} a self-edge is added to a vertex in the state transition graph if the corresponding cubical top cell in $\cX$ is such that \eqref{eq:rate_of_change} has opposite signs at each pair of opposite walls of the cubical cell. This way we can identify cells corresponding to stable and unstable fixed points. Here, in addition to the self-edges described above, we also add a self-edge to a top cell $\sigma \in \cX$ if the following case: If one of the walls of $\sigma$ corresponds to the threshold $0$ then we ignore this wall and its opposite wall when deciding if a self-edge should be added to $\sigma$, that is, we assign a self-edge to $\sigma$ if \eqref{eq:rate_of_change} has opposite signs at each pair of walls that both do not correspond to the threshold $0$. This allows us to identify cells corresponding to fixed points where some of the populations are extinct. Notice that this implies that the top cell at the origin is always identified as a fixed point (which we identify with the extinction equilibria).

\subsection{Multiple edges in DSGRN}

One consequence of allowing multiple edges in DSGRN is that it introduces restrictions on the values of $\Lambda_n(x)$ depending on the relative values of the thresholds. For example, let $(i, j, 0)$ and $(i, j, 1)$ be two edges from node $i$ to node $j$ and assume that one of the terms of $\Lambda_n(x)$ is $\sigma^{(0)}_{j, i}(x_i) + \sigma^{(1)}_{j, i}(x_i)$. If $\theta^{(0)}_{j, i} < \theta^{(1)}_{j, i}$ and the variable $x_i$  associated  with node $i$ satisfies $x_i < \theta^{(0)}_{j, i}$, then we also have that $x_i < \theta^{(1)}_{j, i}$ and so we must have that $\sigma^{(0)}_{j, i}(x_i) = 0$ and $\sigma^{(1)}_{j, i}(x_i) = 0$. Hence the value of $\Lambda_n(x)$ corresponding to $\sigma^{(0)}_{j, i}(x_i) = 0$ and $\sigma^{(1)}_{j, i}(x_i) = \delta^{(1)}_{j, i}$, that is, corresponding to the term $\sigma^{(0)}_{j, i}(x_i) + \sigma^{(1)}_{j, i}(x_i) = \delta^{(1)}_{j, i}$, is never attained. However, these restrictions do not interfere with the parameter decomposition, and hence computing the parameter decomposition to a multi-edge network can be done as described above.

When computing the STG for a given parameter in the parameter graph (which implies a fixed order of thresholds), the location in phase space corresponding to each node of the STG gives us the correct value for $\Lambda_n(x)$ for that node, and hence the restrictions on the values of $\Lambda_n(x)$ are automatically resolved when computing the STG. Once the STG is constructed, the remaining computations are the same as in the original DSGRN.

\begin{remark}
The restrictions on the values of $\Lambda_n(x)$ described above cause some parameter regions to have the same STG (which does not occur in the original DSGRN). We will denote the low values of $\sigma^{(0)}_{j, i}(x_i)$ and $\sigma^{(1)}_{j, i}(x_i)$ by $\ell^{(0)}_{j, i}$ and $\ell^{(1)}_{j, i}$ instead of $0$ to distinguish between them. Then if, for example, the difference between two parameter nodes is given by $\ell^{(0)}_{j, i} < \theta^{(k)}_{\ell, j} < \ell^{(1)}_{j, i} + \delta^{(1)}_{j, i}$ and $\ell^{(0)}_{j, i} < \ell^{(1)}_{j, i} + \delta^{(1)}_{j, i} < \theta^{(k)}_{\ell, j}$, then under the condition above that $\theta^{(0)}_{j, i} < \theta^{(1)}_{j, i}$, these two parameter nodes will produce the same STG since the values $\sigma^{(0)}_{j, i}(x_i) = \ell^{(0)}_{j, i}$ and $\sigma^{(1)}_{j, i}(x_i) = \ell^{(1)}_{j, i} + \delta^{(1)}_{j, i}$ cannot happen at the same time (in the same domain in the STG) due to the restriction $\theta^{(0)}_{j, i} < \theta^{(1)}_{j, i}$ and hence the difference between these two parameters have no effect in the STG for this particular order of the thresholds $\theta^{(0)}_{j, i}$, $\theta^{(1)}_{j, i}$.
\end{remark}

\subsection{Simplified network diagrams}

In all the networks considered in this paper, for every edge $(i, j)$ with $i \neq j$ there is also a corresponding edge $(j, j)$ with the same sign as the edge $(i, j)$ and both edges $(i, j)$ and $(j, j)$ are combined into a term of size $2$ in the interaction function. Since this is true for all the networks considered in the computations in this paper, we opted to present a simplified version of the network diagrams where we omitted the additional self-edges $(j, j)$.

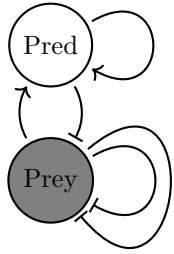
\begin{figure}[!htbp]
\centering
\begin{picture}(60,95)
\put(0,-30){
\begin{tikzpicture}[thick, scale=0.9]
\node[circle, fill=white, draw] (1) at (0,2) {{\small Pred}};
\node[circle, fill=gray, draw] (2) at (0,0) {{\small Prey}};
\draw[-|, shorten <= 2pt, shorten >= 2pt] (1) to [bend left] (2);
\draw[->, shorten <= 2pt, shorten >= 2pt] (2) to [bend left] (1);
\draw[->, shorten <= 2pt, shorten >= 2pt] (1) to [out=30, in=-30, looseness=6] (1);
\draw[-|, shorten <= 2pt, shorten >= 2pt] (2) to [out=30, in=-30, looseness=6] (2);
\draw[-|, shorten <= 2pt, shorten >= 1pt] (2) to [out=40, in=-50, looseness=7] (2);
\end{tikzpicture}
}
\end{picture}
\caption{Network used for the DSGRN computations corresponding to the simplified network in Figure~\ref{fig:2species}A.}
\label{fig:pred_prey_DSGRN_network}
\end{figure}

For example, the network in Figure~\ref{fig:2species}A is the simplified version, and corresponds to the DSGRN network in Figure~\ref{fig:pred_prey_DSGRN_network}. Denoting $x_1$ the variable corresponding to the prey and $x_2$ the variable corresponding to the predator, the interaction function for the prey and predator are $-x_1 - x_1 x_2$ and $x_1 x_2$, respectively. Their rates of change, as given in Section~\ref{sec:2species}, are
\[
\begin{cases}
~~~\gamma_1 x_1 - \sigma^{(0)}_{1,1}(x_1) - \sigma^{(1)}_{1,1}(x_1) \sigma^{(0)}_{1,2}(x_2) & \\
-\gamma_2 x_2 + \sigma^{(0)}_{2,1}(x_1) \sigma^{(0)}_{2,2}(x_2). &
\end{cases}
\]

\newpage
\section*{Acknowledgements}
The work of W.S.C. was supported by the National Science Foundation under award DMS-2138085. W.S.C. was also partially supported by the Laboratory Directed Research and Development (LDRD) Program at Los Alamos National Laboratory. J.A.B. was supported by the National Science Foundation under award DMS-2052616. J.A.B. was also supported by the Simons Foundation under award No. $82610$. The work of K.M. and M.G. was partially supported by the National Science Foundation under awards DMS-1839294 and HDR TRIPODS award CCF-1934924, DARPA contract HR0011-16-2-0033,  National Institutes of Health award R01 GM126555, and Air Force Office of Scientific Research under award numbers FA9550-23-1-0011 and AWD00010853-MOD002. M.G. was also supported by CNPq grant 309073/2019-7. K.M. was also supported by a grant from the Simons Foundation.

\clearpage
\newpage

\bibliography{references}

\end{document}